\documentclass[twocolumn,showpacs,preprintnumbers,amsmath,amssymb,pre]{revtex4-1}

\usepackage{color}    
\usepackage{graphicx}
\usepackage{dcolumn}
\usepackage{bm}
\usepackage{subfigure}
\usepackage{amssymb}
\usepackage{multirow}
\graphicspath{{plots/}}

\renewcommand{\vec}[1]{\mathbf{#1}}

\begin{document}


\title{The ion potential in warm dense matter: \\wake effects due to streaming degenerate electrons}


\author{Zhandos Moldabekov$^{1,2}$, Patrick Ludwig$^1$, Michael Bonitz$^1$, Tlekkabul Ramazanov$^2$}

\affiliation{
 $^1$Institut f\"ur Theoretische Physik und Astrophysik, Christian-Albrechts-Universit\"at zu Kiel,
 Leibnizstra{\ss}e 15, 24098 Kiel, Germany}

\affiliation{
$^2$Institute for Experimental and Theoretical Physics, Al-Farabi Kazakh National University, 71 Al-Farabi str.,  
 050040 Almaty, Kazakhstan
}


\begin{abstract}
The effective dynamically screened potential of a classical ion in a stationary flowing quantum plasma at finite temperature is investigated. This is a key quantity for thermodynamics and transport of dense plasmas in the warm dense matter regime. This potential has been studied before within hydrodynamic approaches or based on the zero temperature Lindhard dielectric function. Here we extend the kinetic analysis by including the effects of finite temperature and of collisions based on the Mermin dielectric function. The resulting ion potential exhibits an oscillatory structure with attractive minima (wakes) and, thus, strongly deviates from the static Yukawa potential of equilibrium plasmas. 
This potential is analyzed in detail for high-density plasmas with values of the Brueckner parameter in the range $0.1 \le r_s \le 1$, for a broad range of plasma temperature and electron streaming velocity. It is shown that wake effects become weaker with increasing temperature of the electrons. 
Finally, we obtain the minimal electron streaming velocity for which attraction between ions occurs. This velocity   
turns out to be less than the electron Fermi velocity. 
Our results allow, for the first time, for reliable predictions of the strength of wake effects in nonequilibrium quantum plasmas with fast streaming electrons showing that these effects are crucial for transport under warm dense matter conditions, in particular for laser-matter interaction, electron-ion temperature equilibration and for stopping power. 
\end{abstract}

\maketitle

\section{Introduction}
Dense plasmas have recently gained growing interest due to their relevance for the interior of giant planets as well as for laser interaction with matter and inertial confinement fusion scenarios. Examples of recent experimental studies include the ultrafast thermalization of laser plasmas \cite{kluge_pop14} or free electron laser excited plasmas \cite{zastrau_prl14}, inertial confinement fusion experiments at the National Ignition Facility \cite{hurricane_nat14} and magnetized Z-pinch experiments at Sandia \cite{cuneo_ieee12, gomez_14}. Questions of fundamental theoretical importance are the energy loss of energetic particles (stopping poser) in such a plasma, e.g. \cite{grabowski_prl13} or the temperature equilibration of the electronic and ionic components \cite{zastrau_prl14,benedict_pre12}. 

Despite recent advances in modeling and computer simulations a fully selfconsistent treatment of these highly nonequilibrium electron-ion plasmas has not been possible so far due to the requirement of electronic quantum and spin effects together with the (possibly) strong ionic correlations. The main problem here are the vastly different time scales of electrons and ions resulting from their different masses.
A possible solution of this dilemma is a multi-scale approach that has been proposed by Ludwig {\em et al.} in Ref. \cite{Patrick2}. It takes advantage of the weak electron-ion coupling that allows for a linear response treatment of the elctrons. This idea has been used by Graziani {\em et al.} to decouple the electron kinetic equation using an STLS (Singwi-Tosi-Land-Sj\"olander) scheme \cite{graziani}. Further improvements along this line should be possible with an extension of STLS recently proposed by K\"ahlert {\em et al.} \cite{kaehlert_pre14}.

The key of this multiscale approach is to absorb the fast electron kinetics into an effective screened potential $\Phi$ of the heavy ions with charge $Q_i$ where the screening is provided by the electrons via a proper dielectric function $\epsilon$, e.g. \cite{Patrick2}
\begin{equation} \label{POT_stat}
\Phi(\vec r)   = \int\!\frac{\mathrm{d}^3k}{2 \pi^2 } \frac{Q_{i}}{k^2 \epsilon(\vec k, 0)} e^{i \vec k \cdot \vec r} \quad,
\end{equation}
taken in the static limit, $\omega=0$. This is justified when the electrons are in (or close to) thermodynamic equilibrium where the potential reduces to the familiar statically screened Yukawa or Debye potential. In a second step of the multiscale approach, the dynamics of the ions is computed exactly by using molecular dynamics simulations involving the screened potential $\Phi$ \cite{Patrick2,graziani}. 
However, under nonequilibrium conditions of fast directed motion of electrons with respect to the ions (electron beams, fast ion stopping in quantum plasmas and metals, laser acceleration of electrons etc.) this approximation brakes down, and generalizion of the potential (\ref{POT_stat}) to the case of a dynamic dielectric function is necessary. 

It is the purpose of this paper to present this generalization for situations relevant to warm dense matter and obtain results that are quantitatively reliable allowing for predictions that can be tested in experiments. To this end we will use a dielectric function of quantum degenerate electrons 
streaming with a constant velocity ${\bf u}_e$ relative to the ions that fully includes the effects of finite temperature $T$ and  collisions (correlations). We will use the Mermin dielectric function \cite{Mermin} and obtain the  generalization of  Eq.~(\ref{POT_stat}) to a dynamically screened ion potential 
\begin{equation} \label{POT}
\Phi(\vec r)   = \int\!\frac{\mathrm{d}^3k}{2 \pi^2 } \frac{Q_{i}}{k^2 \epsilon(\vec k, \vec k \cdot \vec u_e)} e^{i \vec k \cdot \vec r} \quad.
\end{equation}
This potential may radically differ from a Debye potential in a finite range of streaming velocities $u_e$ and even change sign (wake effects). This gives rise to an attraction between two equally charged ions which may significantly influence the static and dynamic properties of dense plasmas subject to streaming particles. 

Such wake effects that are due to a ``focusing'' of the light particles behind the heavy one are well known from classical dusty (complex) plasmas where ion focusing  behind a highly charged heavy dust particle is well established experimentally, e.g. \cite{Tsytovich,Ramazanov1,block12}. 
Recently, it has been shown that the presence of a strong ion drift leads to the excitation of ion instabilities and the destabilization of highly ordered dust grain ensembles such as dust crystals and to their melting \cite{Patrick1}, and also generalizations to magnetized plasmas have been performed \cite{joost_ppcf14}. 
The theoretical concepts to compute the dynamically screened potential are similar to the ones used in this paper and are based on a classical dielectric function derived either from cold fluid theory or from kinetic theory with collisions \cite{Patrick4,Patrick3}. An important outcome of these studies is that the dynamically screened potential derived from linear response theory agrees very well with full nonlinear kinetic simulations, e.g. \cite{Patrick3,block12}.

A second example of wake effects are ion beam experiments \cite{Malka} where the dynamics of ions penetrating a plasma are strongly influenced by wake effects. A third realization of wake effects is observed in condensed matter systems. In fact, the so-called vicinage effect (force) \cite{Echenique,Barriga1} reported in experiments on the energy loss of charged particles in solids is nothing but a wake effect. 

While the dynamically screened potential in a streaming classical plasma is well studied, the corresponding problem in a quantum plasma is much poorer understood theoretically.
A numerical approach to the dynamics of an ion penetrating into a solid is given by quantum-classical Ehrenfest dynamics,   which, however, is very time-consuming, e.g. \cite{Race,Mao}.
On the other hand, the interaction of a fast ion with the electrons of the solid and the related stopping power were investigated in detail using a dielectric approach, similar to our concept. The dielectric function of the target material was computed e.g. in Refs.~\cite{Abril,Barriga2,Barriga3,Schin}, however, only in the zero-temperature limit. A similar analysis involving 
the zero-temperature dielectric function of streaming quantum electrons and neglecting electron-electron collisions was performed in Ref.~\cite{Else}
where the behavior of the Friedel oscillations was studied.

Further, we mention a classical molecular dynamics approach to charged-particle stopping in warm dense matter by Grabowski {\em et al.}~\cite{grabowski_prl13} where the classical wake potential was computed. Finally, an even simpler approach that is based on a quantum hydrodynamic model (QHD) has been applied that predicts an attractive interaction between ions even in the absence of streaming, $u_e = 0$ \cite{shukla_prl12}. Comparisons with density functional theory revealed that this is incorrect \cite{bonitz_pre13} and points to limitations of QHD models for dense plasmas \cite{bonitz_psc13,krishnaswami14}.
A recent overview and more references can be found in Ref.~\cite{Vladimirov}. 

While many of the above works indicated the principal importance of wake effects in dense quantum plasmas with fast projectiles, particle beams or streaming electrons, 
an analysis  that allows for reliable quantitativ predictions under conditions relevant to warm dense matter is still missing.  
The present paper aims at filling this gap. To this end we perform calculations of the dynamically screened potential of an ion in the presence of streaming quantum electrons fully including finite temperature effects and electron-electron collisions thereby scanning a broad range of densities, temperatures and streaming velocities.
The present kinetic treatment yields results that are substantially different both from classical wake potentials \cite{Patrick3} and from quantum potentials obtained within quantum hydrodynamics \cite{michta_wakes14}. This indicates that the QHD approach should be applied to warm dense matter with great caution as it may lead to wrong results.

The paper is organized as follows: After discussing the relevant parameters (section \ref{s:2}) 
we introduce the Mermin dielectric function in Sec.~\ref{s:mermin} and discuss the chosen model for the electron-electron collision frequency. The dynamically screened potential is presented in Sec.~\ref{s:potential} and analyzed for a broad range of plasma parameters. Finally  in Sec~\ref{s:dis} we present a detailed discussion of our results and of the limitations of our model. The paper concludes with an Appendix that contains details of the evaluation of the quantum dielectric function as well as additional results for an ultra-relativistic quantum plasma. These results indicate analogous quantum plasma behavior in the cases of weak and strong relativistic effects.

\section{Considered density and temperature range. Dimensionless parameters} \label{s:2}
In this paper we study a dense low-temperature plasma containing classical--possibly strongly correlated--ions embedded into a streaming quantum electron plasma. Such a two-component plasma (for simplicity, in the simulations below, we assume equal temperatures of electrons and ions) is characterized by the following energy scales \cite{bonitz_pop08}: thermal energy (per particle), $\frac{3}{2} k_BT$, the electron Fermi energy 
$E_F=mv_F^2/(2m) \equiv k_BT_F$, where $v_F$ and $T_F$ are the Fermi velocity and Fermi temperature, respectively,
and the mean electron streaming engergy will be denoted by $E_U=u_e^2/(2m)$. Further, the interaction energy of free electrons is characterized by the plasmon energy, $\hbar \omega_p$ with the plasma frequency $\omega_p=(4\pi n e^2/m)^{1/2}$, whereas the scale for bound electrons is the atomic ground state binding energy, $E_R = Ze^2/2a_B$.  The relevant length scales are 
the mean interparticle distance of the electrons (ions) $a$ ($a_i$) and the Bohr radius $a_B$. In this paper we will consider a hydrogen plasma (the results are directly applicable to multiply charged ions by a rescaling of the potential, see below), where $a_B$ is the hydrogen Bohr radius and the binding energy is related to the familiar Hartree energy by $E_R=13.6 {\rm eV}={\rm Ha}/2$.
In the following, we will use atomic units (a.u.), where $m_{e}=e=\hbar=1$, i.e. lengths are given in units of the Bohr radius and energies in units of Hartree. 

The plasma state is conveniently characterized by the following dimensionless parameters: 
\begin{description}
 \item[i.] the electron degeneracy parameter, $\theta=k_{B}T/E_{F}$,
 \item[ii.] the quantum coupling parameter (Brueckner parameter), $r_{s}=a/a_{B}$.
[an alternative parameter is $\eta=\hbar \omega_{p}/4E_{F}$ which is related to the Brueckner parameter by  $\eta\simeq\sqrt{r_{s}/18.1}$],
 \item[iii.] the coupling parameter of the ions, $\Gamma_i = Q_i^{2}/(a_i k_BT)$, where the ion charge is $Q_i=-Ze$,
 \item[iv.] the dimensionless streaming velocity (Mach number) is defined as $M=u_{e}/v_{F}$.          
\end{description}

Relativistic effects are not important for typical warm dense matter conditions and will, therefore, be discarded in the following [for a discussion 
of wake effects in ultra-relativistic plasmas, see Appendix B].  
 This puts a lower bound to the Brueckner parameter, $r_{s} \gg 0.014$, that follows from the condition $v_F \ll c$.
In the calculations below we will restrict ourselves to $r_s \ge 0.1$. 

Thus the plasma in equilibrium is characterized by two parameters: $\theta$ and $r_{s}$. 
The temperature in Kelvin or eV then follows according to 
$T\simeq\frac{\theta}{r_{s}^2}\cdot 0.58 \times  10^6 \left[ K\right]$
and 
$k_{B}T\simeq\frac{\theta}{r_{s}^2}\cdot 50.12 \left[eV\right]$.
The range of parameters considered in the present work is set by the applicability limits of the theory, which assumes 
weak electron coupling ($r_s \lesssim 1$) as well as weak electron-ion coupling which is the basis for the linear response ansatz for the screened ion potential. At the same time the ion coupling can be strong ($\Gamma_i \gg 1$), and can be studied, e.g. by performing molecular dynamics simulations with the dynamically screened ion potentials derived in the present paper \cite{Patrick2}.
Typical density and temperature parameters used below are listed in Table~\ref{t:parameters}, for the case $\theta=1$ (i.e. $k_BT=E_F$) and are trivially rescaled to other values of $\Theta$.

In the present nonequilibrium plasma case of streaming electrons, the plasma is characterized by a third dimensionless parameter---the Mach number $M$.	

\begin{widetext}
\begin{table}[h]
\caption{Examples of plasma parameters used in this paper (numbers refer to the case $\theta=1.0$).}
\begin{center}
\begin{tabular}{|c||*{8}{c|}c}
\hline
$r_s$              & 0.1 & 0.2 & 0.3 & 0.5 & 0.8 & 1 & 1.5 & 4.52 \\
\hline\hline
$n ({\rm cm}^{-3})$	   & $1.61\times 10^{27}$& $2.0\times 10^{26}$& $6.0\times 10^{25}$& $1.3\times 10^{25}$& $3.1\times 10^{24}$& $1.6\times 10^{24}$ & $4.74 \times 10^{23}$ & $1.73 \times 10^{22}$\\  
\hline
$T (K)$		   & $5.8\times 10^7 $& $1.45\times 10^7 $& $6.44\times 10^6 $& $2.32\times 10^6 $& $9\times 10^5 $& $5.8\times 10^5 $ & $2.58\times 10^5$ & $0.28\times 10^5$\\
\hline
$k_{B}T ({\rm eV})$	   & $5\times 10^3 $& $1.2\times 10^3 $& $5.57\times 10^2 $& $2\times 10^2 $& $78 $& $50 $ & $22.27$& $2.45$\\
\hline
$k_{B}T ({\rm a.u.})$	   & $184 $& $46 $& $20.44 $& $7.36$& $2.87 $& $1.84$ & $0.82$& $0.09$\\
\hline
\end{tabular}
\end{center}
\label{t:parameters}
\end{table}
\end{widetext}

As mentioned in the introduction, the largely different time scales of electrons and ions prevent a selfconsistent time-depedent quantum simulation. Instead we will apply our multiscale approach \cite{Patrick2} where the whole information on the electron component is condensed in a dynamic dielectric function which determines the effective potential of a single ion.
Thus, the quality of the results depends on the accuracy of the dielectric function. While there exists an extensive literature on finite temperature dielectric functions with correlation effects, e.g. \cite{kwong_prl_00}, many results are very complicated and difficult to implement in plasma simulations. Therefore, 
here we will use the particle number conserving relaxation time approximation due to Mermin \cite{Mermin} with a temperature and density dependent collision frequency, which has not been applied before to the effective ion potential in warm dense matter.
%

\section{Mermin dielectric function, collision frequency and perturbed electron density}\label{s:mermin}
The standard mean-field (Hartree or quantum Vlasov) result for the electron dielectric function is given by the random phase approximation (RPA), $\epsilon_{RPA}(k,\omega)$, where correlation effects are neglected, and damping of collective oscillations is entirely due to Landau damping. In the warm dense matter regime we expect the electrons to be weakly to moderately coupled so that correlation effects are, in general, relevant and approximations beyond the RPA are often required. 

The simplest quantum dielectric function which takes collisions into account in a conserving fashion \cite{df_damping}
was derived by  Mermin \cite{Mermin} in relaxation time approximation and has the form:
\begin{multline}\label{Mermin}
 \epsilon_{M}(k,\omega)=1+\\ +\frac{(\omega+i\nu)[\epsilon_{RPA}(k,\omega+i\nu)-1]}{\omega+i\nu[\epsilon_{RPA}(k,\omega+i\nu)-1]/[\epsilon_{RPA}(k,0)-1]},
\end{multline}
where $\nu$ is the electron collision frequency.
\begin{figure}[t]
\includegraphics[width=0.45\textwidth]{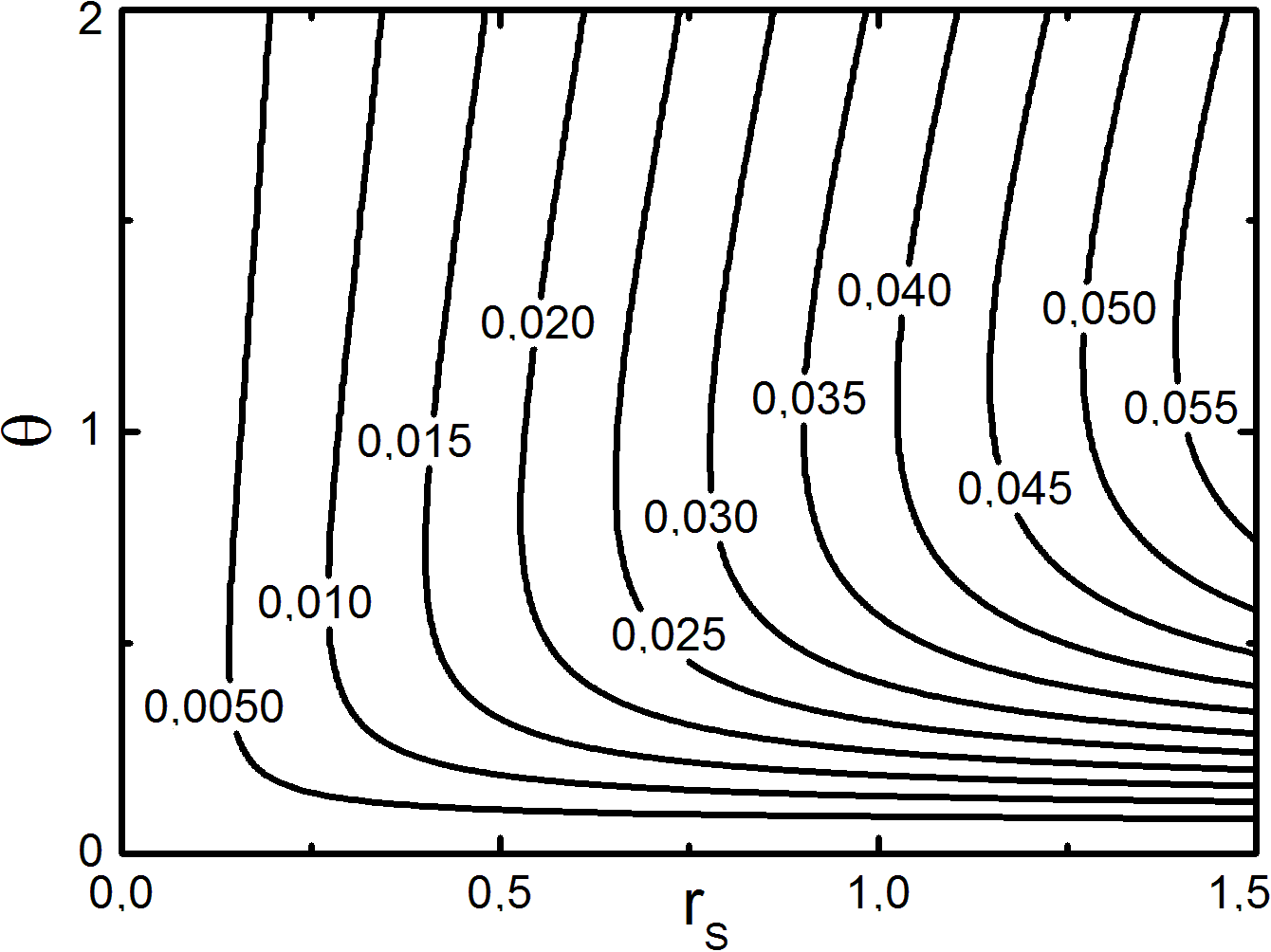}
\caption{Contours of constant electron-electron collision frequency $\nu$, Eq.~(\ref{ nu1 }), in units of $\tilde{\omega}_{p}$.}
\label{fig:coll}
\end{figure}
The dielectric function (\ref{Mermin}) involves the RPA dielectric function for a finite temperature which is summarized in the Appendix A. 
A derivation of the Mermin dielectric function from quantum kinetic theory  was given in \cite{Patrick2}.

The dielectric function (\ref{Mermin}) contains an energy-independent electron-electron collision frequency, $\nu = \nu(n, T)$. A simple parametrization 
valid for arbitrary degeneracy, has been given in Ref. \cite{Barriga2} and will be used below
\begin{equation}\label{ nu1 }
 \nu=\frac{\nu_{0}}{\sqrt{1+0.2T/T_{F}}},
\end{equation}
where $\nu_{0}=\nu(k_{B}T \ll E_{F})$, given by
\begin{equation}\label{nu0}
 \nu_{0}=\frac{3(k_{B}T)^2}{2\hbar m_{e} c^2}\sqrt{\frac{\alpha x^3}{\pi^3(1+x^2)^5/2}}\,J(y)\,;
\end{equation}
and $x=v_{F}/c$, $y=\sqrt{3}\hbar \tilde{\omega}_{p}/k_BT$, and $\tilde{\omega}_{p}=[4\pi e^2n_e/(m_{e}(1+x^2)]^{1/2}$.
The function $J(y)$ has the form~\cite{Potekhin}
\begin{equation}\label{ J }
\begin{split}
 J(y)&= \bigg[  \frac{y^3}{3(1+0.07414y)^3}\times\ln\left(\frac{2.810}{y}-\frac{0.810 x^2}{y(1+x^2)}+1\right)
\\
& \quad +\frac{\pi^5}{6}\frac{y^4}{(13.91+y)^4}\bigg]
\cdot \left( 1+\frac{6}{5x^2}+\frac{2}{5x^4}\right),
\end{split}
\end{equation}
which is applicable for $0.01\leq x\leq100$, corresponding to the density range $1.4\times10^{-4}\leq r_{s}\leq1.46$. 
As mentioned above, for our analysis of warm dense matter situations we restrict $r_s$ to values above $0.1$, so that
relativistic effects are of minor importance ($x\ll 1$).
In Fig.~(\ref{fig:coll}) the values of the electron-electron collision frequency in units of the plasma frequency are shown. 
With increasing degeneracy, the collision frequency decreases due to the Pauli principle.
%
Furthermore, the collision frequency increases with $r_s$ due to the increased Coulomb coupling.

There exists an abundant literature on the computation of the collision frequency. For example, Ref. \cite{thiele_pre08} contains results for warm dense matter conditions in static Born approximation (see also references cited therein). At low densities ($r_s \sim 1.5$), the results substantially exceed the predictions of the analytical parametrization, Eq.~(\ref{ nu1 }), so 
it has to be expected that the latter does not necessarly account for all collision effects. Moreover, other collision mechanisms such as in the presence of other charged or neutral particle species, may also lead to increased collision frequencies.

However, up to now there are no comprehensive data for the collision frequency in the entire warm dense matter region available.
Therefore, in the majority of our numerical simulations below we will use the analytical expression (\ref{ nu1 }). Yet, in order to assess the general effect of collisions on the dynamically screened ion potential we include a separate subsection \ref{ss:collisions} where we use 
values of the collision frequency $\nu$ that are deliberately chosen larger than those of Eq.~(\ref{ nu1 }).  

From the dielectric function all properties of the streaming electrons that are perturbed by a single ion can be directly computed within linear response. For example the perturbation of the electron density follows as (tilde indicates Fourier transformed quantities, and the electron charge is $-e_0$), \cite{Patrick2}
\begin{equation}
 {\tilde n}^{(1)}({\bf k},\omega) = \Pi_M(k,\omega) (-e_0) {\tilde \Phi({\bf k},\omega)},
\end{equation}
where $\Pi_M$ denotes the longitudinal polarization function with $\epsilon_{M}(k,\omega) = 1 - \frac{4\pi e_0^2}{k^2}\Pi_M(k,\omega)$.
Using the result (\ref{POT}) for the dynamically screened potential this expression becomes 
\begin{equation}
 {\tilde n}^{(1)}({\bf k},\omega) = \frac{Q_i}{e_0}\frac{\epsilon_{M}({\bf k},\omega) - 1 }{\epsilon_{M}({\bf k},\omega) },
\label{eq:e-density}
\end{equation}
and the total electron density follows from the Fourier transform of (\ref{eq:e-density}), together with 
the unperturbed density, $n({\bf r},t) = n (r_s) + n^{(1)}({\bf r},t)$.

\section{Numerical results for the dynamically screened ion potential} \label{s:potential}
Having obtained the retarded longitudinal dielectric function we now can compute the dynamically screened ion potential, Eq.~(\ref{POT}). To this end the Fourier transform of the bare Coulomb potential divided by the Mermin dielectric function 
has to be computed which requires a very high accuracy. We used  a computer program previously developed for classical plasmas~\cite{Patrick3} and extended it to the quantum case.
This code allows for high accuracy (using large grids in Fourier space) wake potential calculations. Its accuracy was 
tested against results obtained with a Mathematica implementation of the plasma dispersion function \cite{Mathematica}. A second thorough test of the code was made 
for the case of a classical plasma by comparison with first-principal nonlinear particle in cell simulations \cite{Patrick4, block12}. The agreement turned out to be excellent, except for very slow particles and/or highly charged heavy particles where linear response theory fails. 

Based on this experience we expect that the present linear response approximation will also be adequate for quantum plasmas in the warm dense matter range. 
However, for quantum systems no reliable nonlinear kinetic tests are available yet. Simulations for inhomogeneous quantum plasmas based on quantum kinetic theory or nonequilibrium Green functions are presently only emerging, e.g. \cite{balzer_lnp_13,bonitz_cpp_13,hermanns_prb14, bonitz_cpp14}.

\subsection{Limit of static screening}\label{ss:static}
\begin{figure}[t]
\includegraphics[width=.8\linewidth]{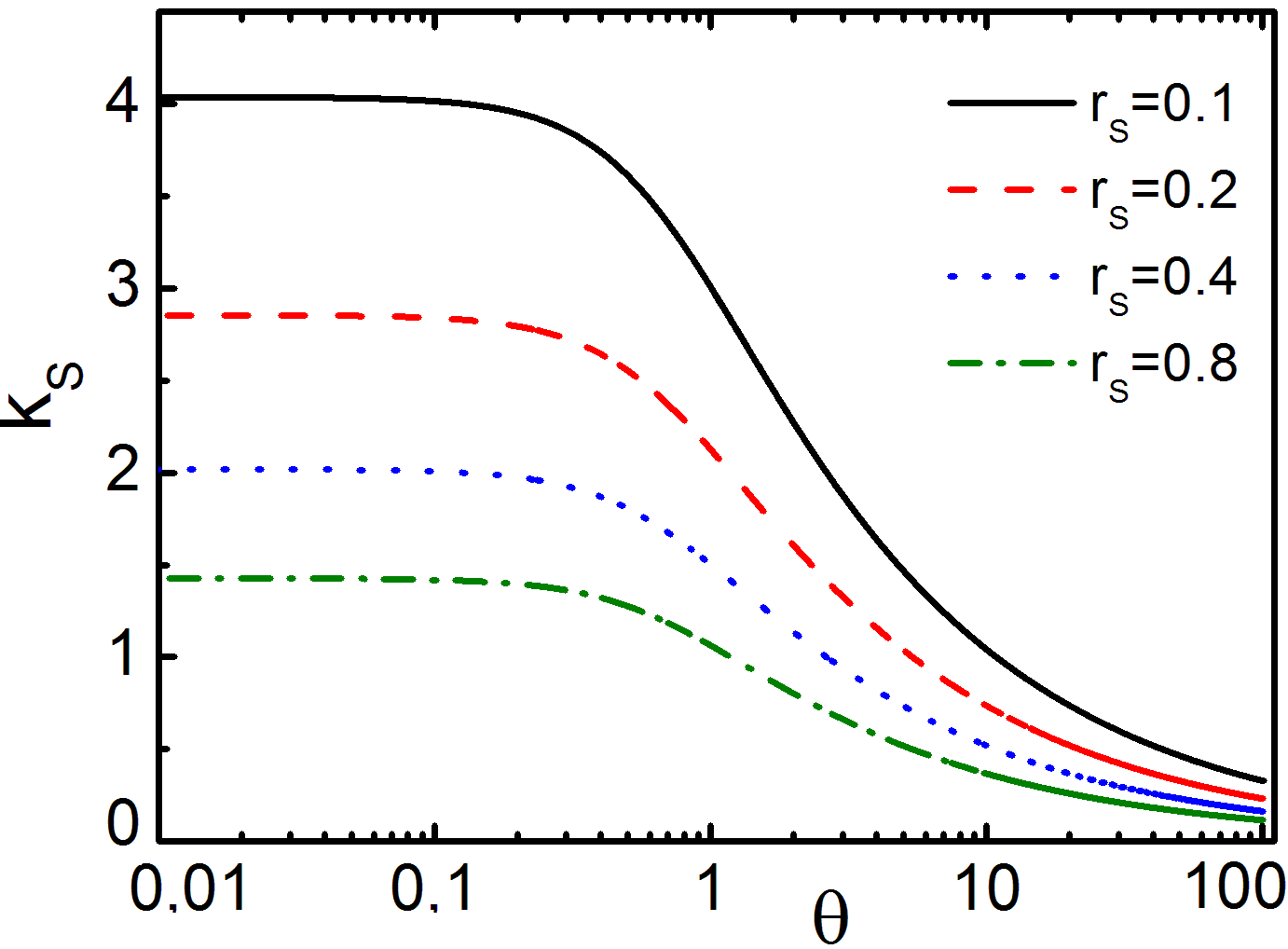}~\hspace{-0mm}\begin{scriptsize}a)\end{scriptsize}
\includegraphics[width=.8\linewidth]{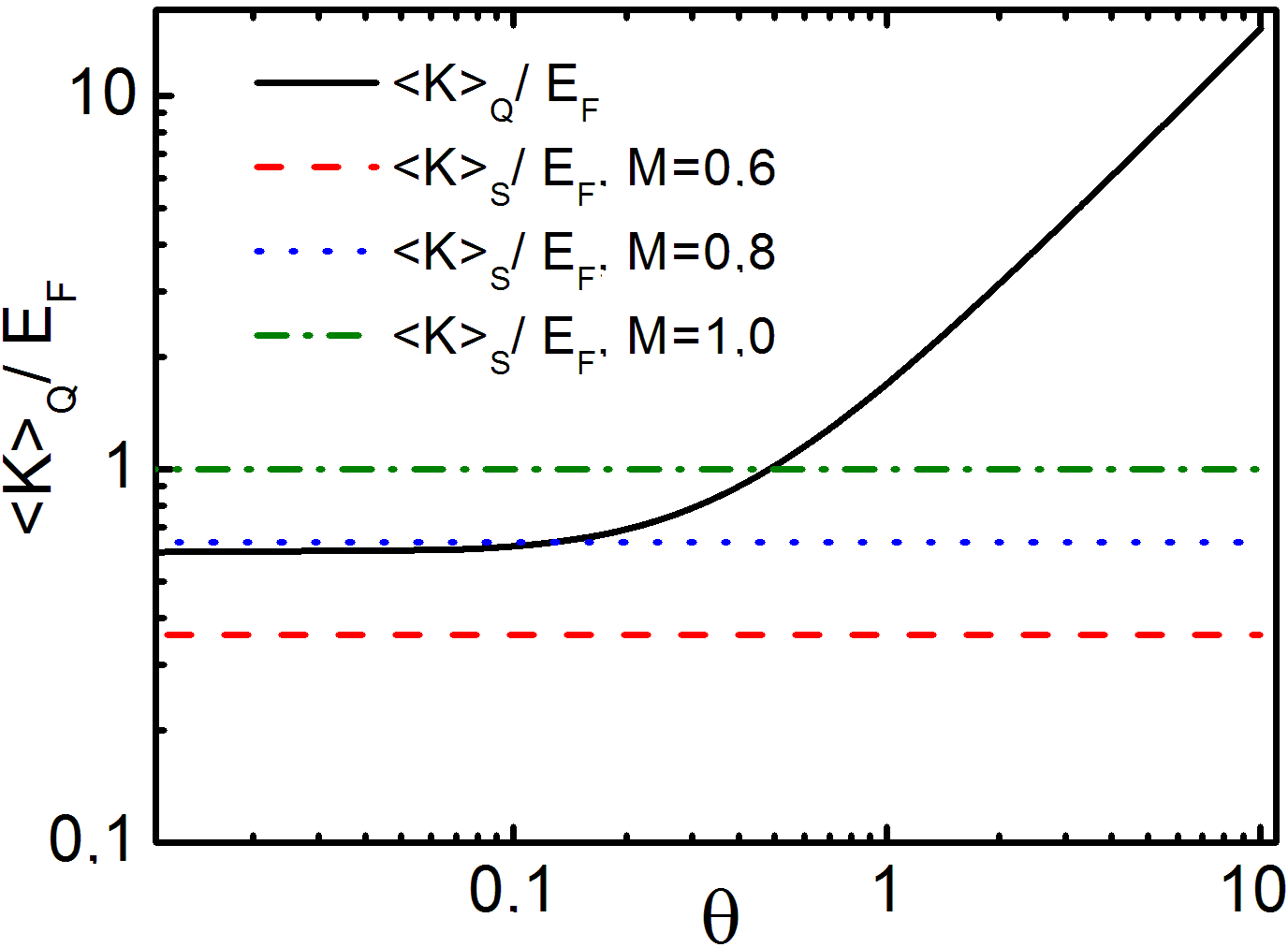}~\hspace{-0mm}\begin{scriptsize}b)\end{scriptsize}
\caption{(Color online) Temperature dependence of static properties. {\bf Top}: Screening parameter $k_{S}$ (in units of $a_B^{-1}$) in the potential (\ref{Yukawa}), for various densities given in the figure. {\bf Bottom}: Quantum kinetic energy $\langle K \rangle_Q $ of electrons in units of the Fermi energy compared to the streaming kinetic energy  $\langle K \rangle_S $ for various streaming velocities. Note that $\langle K \rangle_Q/E_F $ is independent of $r_{S}$, cf. Eq.~(\ref{KQ}).}
\label{fig:ks}
\end{figure}
The first test of our potential (\ref{POT}) is the static collisionless limit ($\nu = 0$) of non-streaming electrons ($u_e \to 0$), where $\Phi$, in linear response, reduces to a Yukawa potential (\ref{POT_stat})
\begin{equation}\label{Yukawa}
\Phi_{Y}=\frac{Q_{i}}{r}\exp (-k_{S}r),
\end{equation}
 where $k_{S}^{-1}$ is the screening length. The screening parameter for an electron gas at finite temperature in RPA \cite{A and B} is
\begin{equation}\label{ ksc }
k_{S}^2=\frac{1}{2} k_{TF}^2 \theta ^{1/2} I_{-1/2}(\beta \mu),
\end{equation}where $k_{TF}=\sqrt{3}\omega_{p}/v_{F}$ is the Thomas-Fermi wave number, and $I_{-1/2}$ is the Fermi integral of order $-1/2$. 

In Fig.~\ref{fig:ks}.a the value of $k_{S}$ is shown as a function of $\theta$ for different values of $r_{s}$. With increasing $r_{s}$ (decreasing density),
the number of electrons in the screening cloud decreases. At the same time, the screening length, $1/k_{S}$, increases. Following a line of given density ($r_s =$ const), 
an increase of $\theta$ is equivalent to an increase of temperature. For $\theta \gg 1$ the classical behavior of the Debye screening length is recovered. In contrast, 
for $\theta < 1$ quantum effects dominate. In particular, at low temperatures with $\theta<0.1$, the screening parameter stays approximately constant. This can be understood by noticing that, at low temperatures, the quantum kinetic energy of the electrons 
\begin{equation}\label{KQ}
\frac {\left< K \right >_{Q}}{E_{F}}=\frac{3}{2}\theta^{5/2}\int_{0}^{\infty}dy\frac{y^{3/2}}{\exp(y-\beta\mu)+1}, 
\end{equation}
also varies slowly with temperature, approaching $3E_F/5$, as can be seen from Fig.~\ref{fig:ks}.b. This figure also provides, for comparison, typical values of the electron streaming kinetic energy, $\left< K\right >_{S}=mu_{e}^{2}/2$  (in units of the Fermi energy) corresponding 
to different streaming velocities $u_{e}=Mv_{F}$.   

\subsection{Friedel Oscillations in a flowing quantum plasma}\label{ss:friedel}
\begin{figure}[t]
\includegraphics[width=0.47\textwidth]{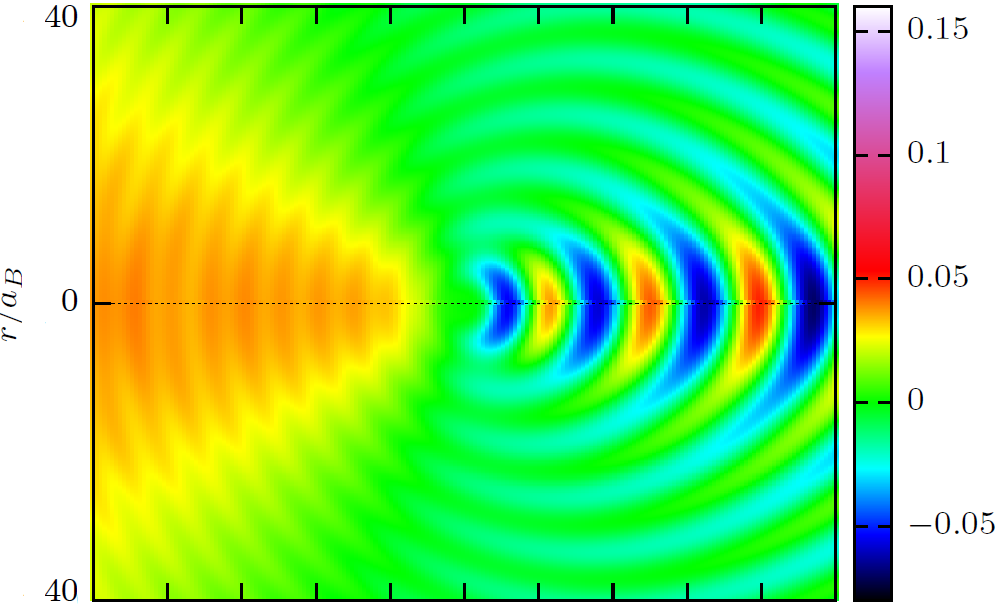}~\hspace{-5mm}\begin{scriptsize}a)\end{scriptsize}
\\
\includegraphics[width=0.455\textwidth]{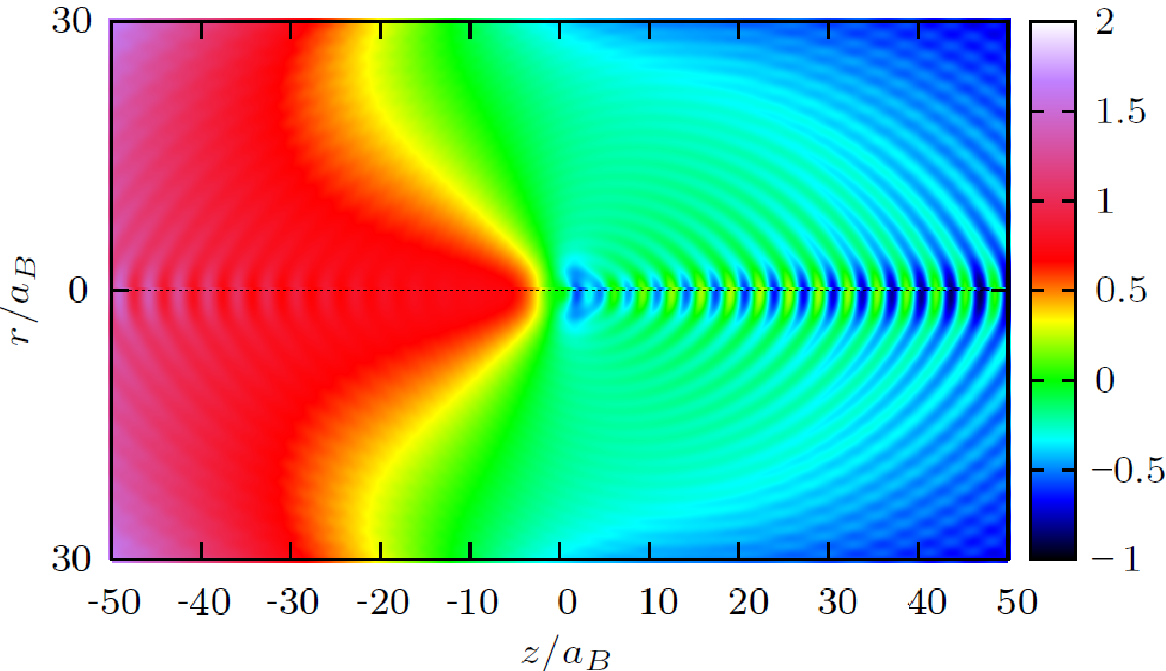}~\hspace{-2.5mm}\begin{scriptsize}b)\end{scriptsize}
\caption{(Color online) Potential $\lvert\vec{r}\rvert^{3} \Phi (\vec{r})$ illustrating the behavior of Friedel oscillations 
in the presence of electrons streaming in positive $z$-direction with $u_{e}=0.5v_{F}$ and $\theta=0.01$. Top figure: $r_{s}=4.52$, bottom: $r_{s}=1$. The result agrees well with Fig. 7 of Ref.~\cite{Else}. Note the strongly increased amplitude of the potential  in the lower figure (different color scale).} 
\label{fig:Fredel}
\end{figure}
Strongly degenerate quantum systems exhibit the so-called Kohn anomaly \cite{Kohn} arising form the step character of the Fermi function at the Fermi surface in momentum space. In coordinate space this translates into long-range oscillatory behavior in the statically screened potential (Friedel oscillations) which is clearly observed e.g. in scanning tunneling spectroscopy experiments of metal surfaces \cite{ziegler_prb_08}. The recovery of Friedel oscillation is an important consistency check for any quantum theory of effective potentials. Friedel oscillations are, for example, not captured by quantum hydrodynamics \cite{bonitz_pre13} or by a Yukawa potential (\ref{Yukawa}). 
Taking the static limit of the potential (\ref{POT}) within the RPA ($\nu \to 0$), for long distances, 
where the Yukawa part of the potential is already damped out, we recover the known asymptotics $\sim\cos (2k_{F}r)/r^3$ \cite{Friedel}. At finite temperature, Friedel oscillations die away from the ion as $\cos (2k_{F}r)/r^2 \exp(-wr)$, where $w=\sqrt{2m}\pi k_{B}T/\sqrt{\mu} \hbar$ \cite{Garssme}. 

Now it is interesting to analyze Friedel oscillations  in the presence of streaming electrons. This was previously investigated  in Ref.~\cite{Else} for a quantum plasma at $T=0$. While these shallow oscillations may be of minor practical relevance for warm dense matter conditions, they provide a useful test of the accuracy of the real-space potential and reflect how well the Fermi statistics are captured.
To visualize the pattern of Friedel oscillations we multiply the full potential by $r^3$ and present the results in Fig.~\ref{fig:Fredel} for the parameters that were used in Ref.~\cite{Else}. We observe very good agreement with 
that reference, although at the lower density ($r_s=4.52$, Fig.~\ref{fig:Fredel}.b) the validity of the present dielectric function (and of the one in Ref.~\cite{Else}) is questionable. Interestingly, at the higher density, $r_s=1$, shown in Fig.~\ref{fig:Fredel}.a the Friedel oscillations are much stronger pronounced (note the different color scale) and the oscillation period is strongly reduced. This is in agreement with the above analytical result for $T=0$ where this period scales with $k^{-1}_F\sim r^{-1}_s$.
%
\subsection{Wake effects in a streaming quantum plasma}\label{ss:wakes}
Let us now turn to the details of the dynamically screened potential, in particular to its dependence on the streaming velocity (Mach number $M$). In the presence of streaming electrons, the formerly istropic screening cloud becomes deformed. 
From a microscopic point of view, this can be understood as the ion ``focusing'' electrons behind itself, thereby creating a spatial region with excess local electron density (probability) which, in turn, may attract other ions. This region will be called ``attractive region'' below.  

While the effect of attraction due focusing of streaming charged particles has been well studied in several fields, in particular in dusty plasmas, e.g. \cite{Patrick1}, a priori it is not known how important such an effect is for ions in warm dense matter under realistic conditions. The present model allows us to give a reliable answer. First of all, the magnitude of the excess electron density depends on a variety of factors, most importantly (a) the streaming velocity $u_{e}$, (b) the density, and (c) temperature (entering via $\theta$). A first overview on the influence of $M, r_s$ and $\Theta$ is presented in Fig.~\ref{fig:M}.
We will now separately analyze the impact of all three factors.
\begin{description}
 \item[a.] {\em Influence of the streaming velocity}.
In a homogeneous equilibrium plasma, the  average number of electrons scattered from an ion is isotropic. In contrast, with the appearence of streaming the number of electrons scattered in flow direction increases, and electrons tend to accumulate behind the ion. Increasing the streaming velocity $u_{e}$ increases the number of scattered electrons, 
thereby enhancing the effect of attraction.  At the same time, it is clear that, beyond a certain value of $u_e$, a further increase will shift the attractive region away from the ion \cite{Mazzaro}. Thus, a weakening of  the attraction should be expected roughly for $M \sim 1$.  This general picture is confirmed by our results shown in Fig.~\ref{fig:M}.a. There we depict the dynamically screened potential $\Phi(r,z)$ in streaming direction $z$ for different values of $M$. Indeed, already for small $M$ a shallow attractive minimum emerges that becomes deeper with increasing $M$ and moves towards the ion. This is explored more in detail below in Fig.~\ref{fig:MinRS}.
\begin{figure}[t]
\includegraphics[width=0.95\linewidth]{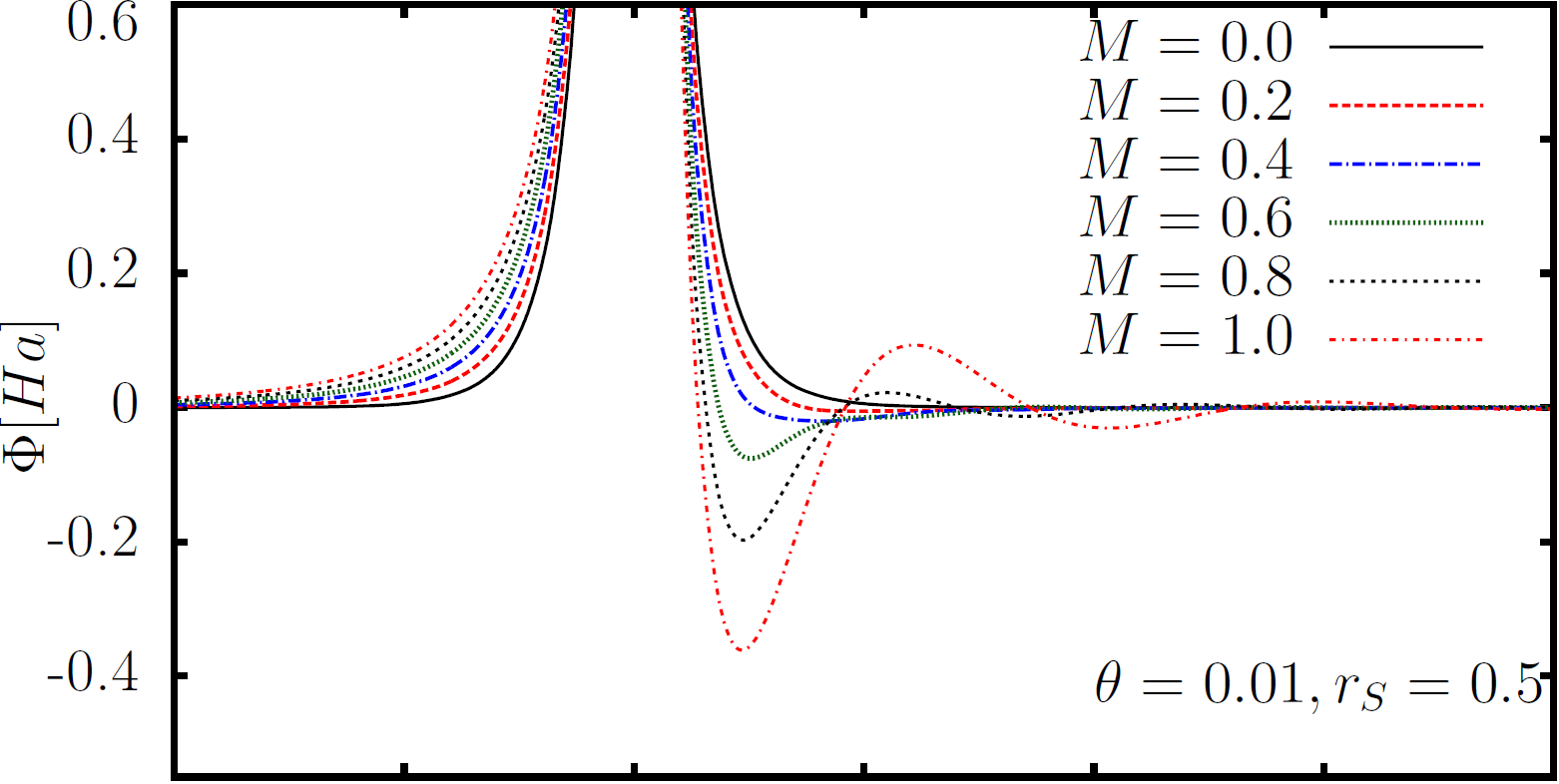}~\hspace{-0mm}\begin{scriptsize}a)\end{scriptsize}
\includegraphics[width=0.95\linewidth]{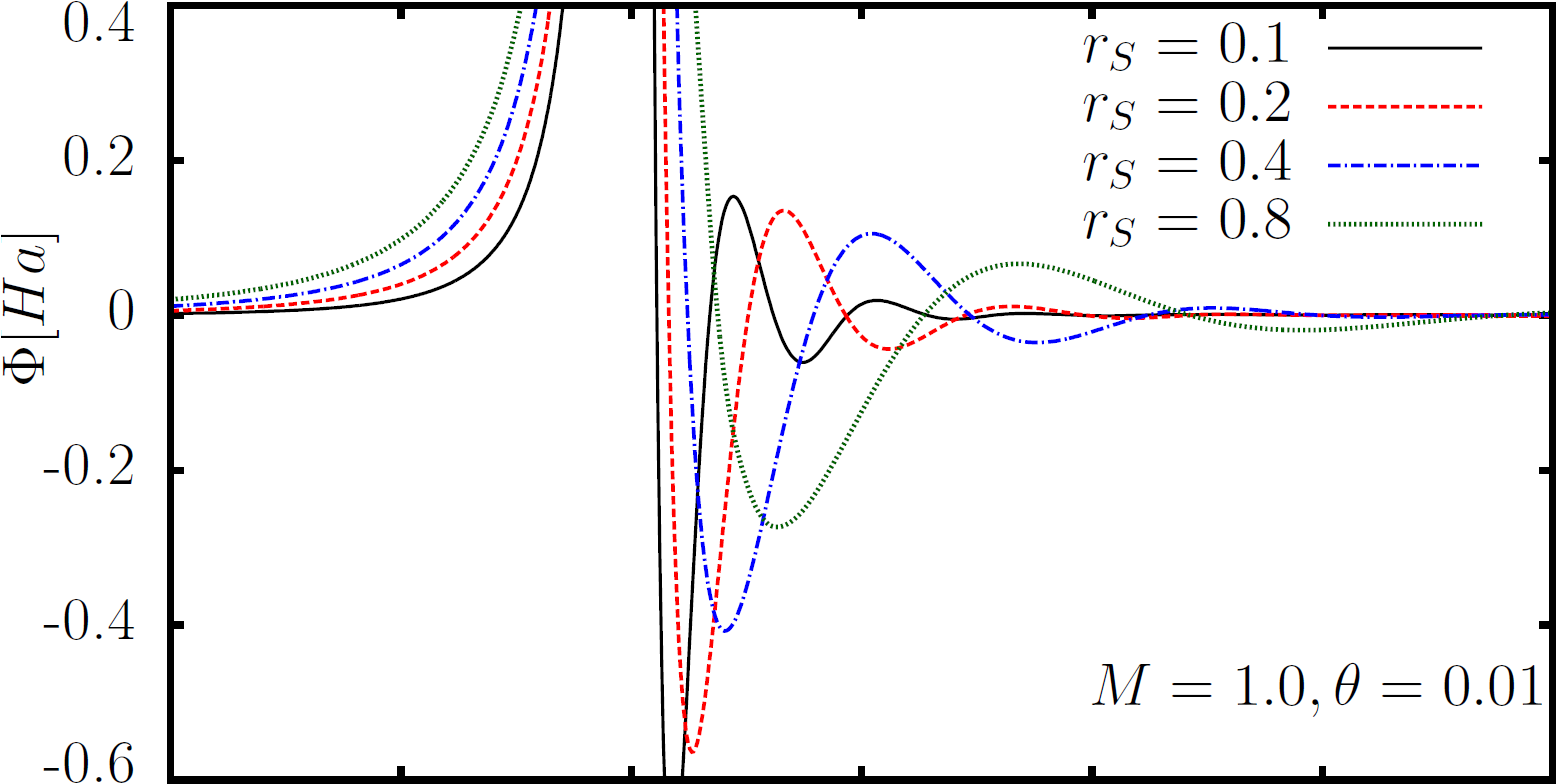}~\hspace{-0mm}\begin{scriptsize}b)\end{scriptsize}
\includegraphics[width=0.95\linewidth]{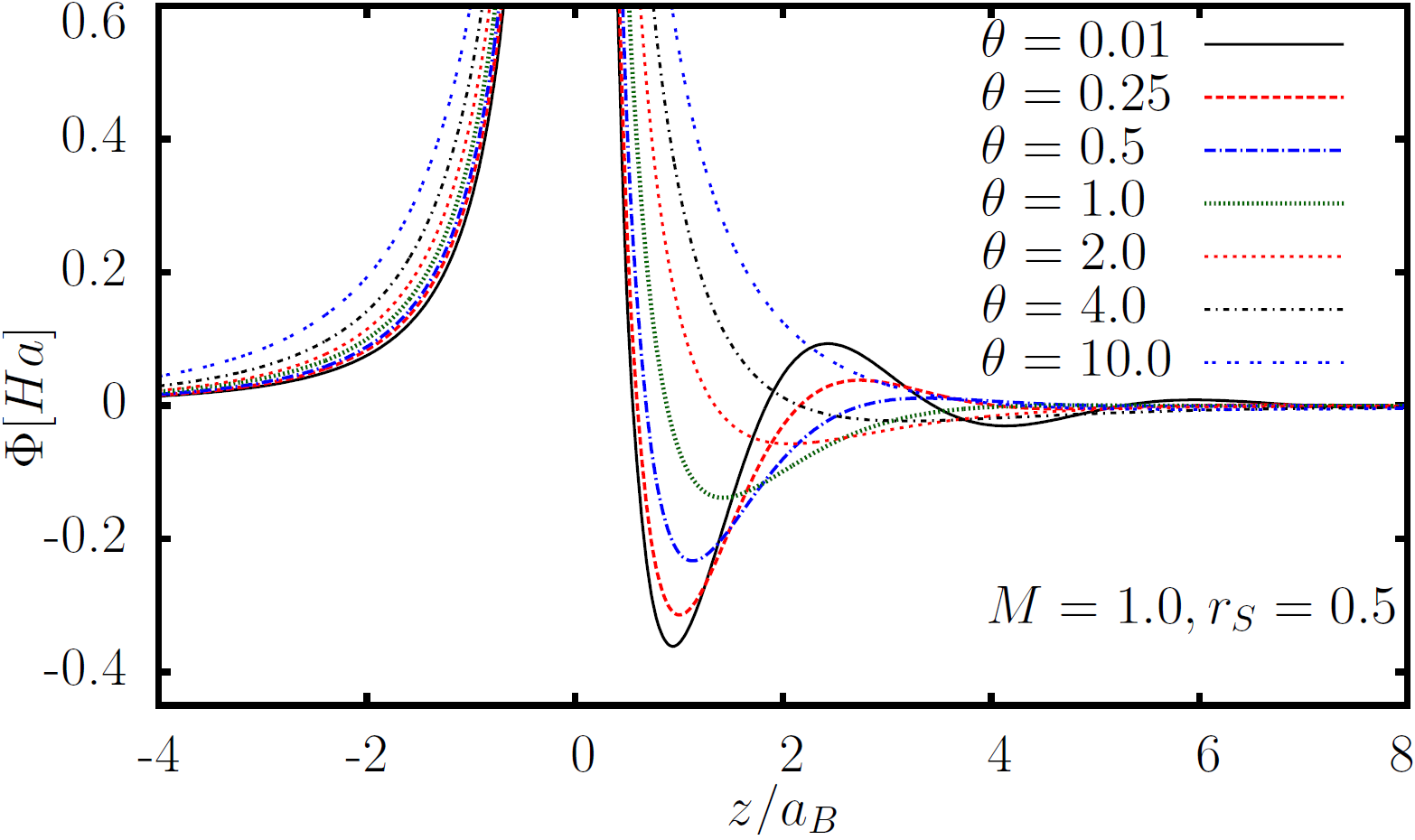}~\hspace{-0mm}\begin{scriptsize}c)\end{scriptsize}
\caption{(Color online) Effective dynamically screened ion potential $\Phi(z)$ and its variation with the streaming velocity $M$ ({\bf top}), density parameter $r_s$ ({\bf center}) and 
temperature $\Theta$ ({\bf bottom}). The potential is shown along the streaming direction $z$ at the ion position ($r=0$).}
\label{fig:M}
\end{figure}
\begin{figure}[h]
\vspace*{0.5cm}
\includegraphics[width=.8\linewidth]{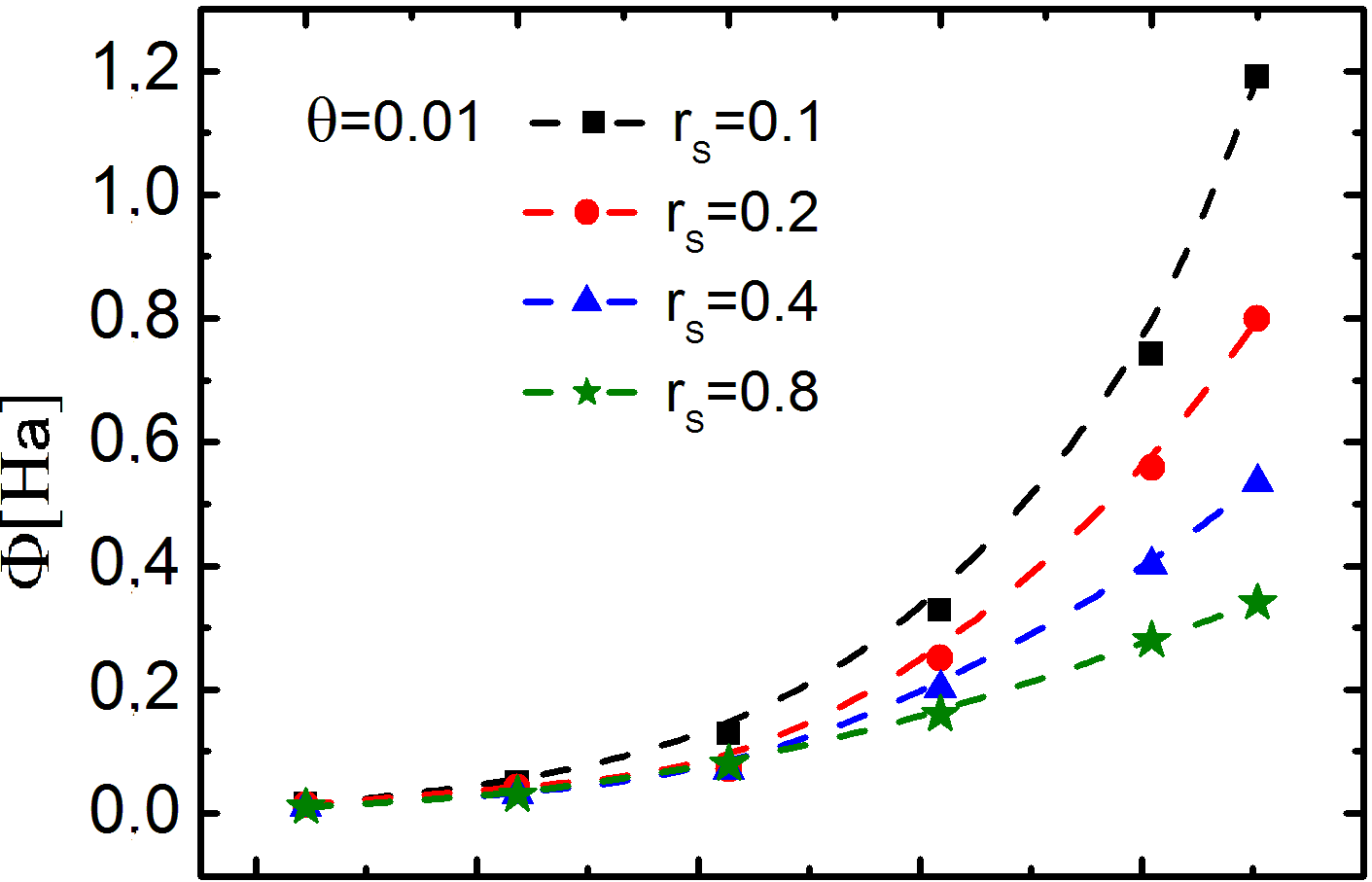}~\hspace{-0mm}\begin{scriptsize}a)\end{scriptsize}
\includegraphics[width=.8\linewidth]{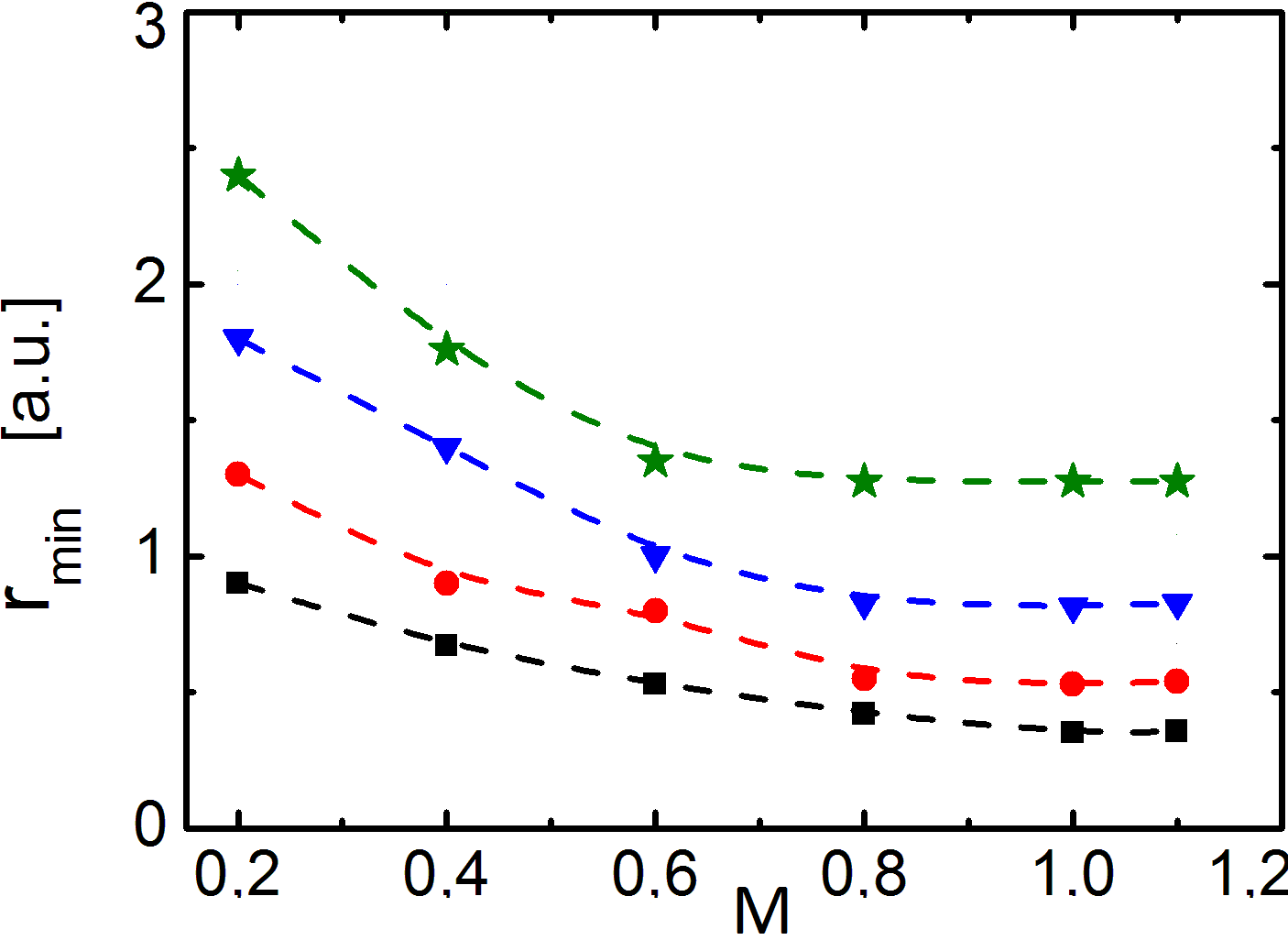}~\hspace{-0mm}\begin{scriptsize}b)\end{scriptsize}
\caption{(Color online) Absolute depth (top) and location (bottom) of the first minimum of $\Phi$ behind the ion  versus streaming parameter $M$ for $\theta=0.01$ and different density parameters $r_{s}$ shown in the figure.}
\label{fig:MinRS}
\end{figure}
\item[b.] {\em Influence of the electron density}. In Fig.~\ref{fig:M}.b the ion potential $\Phi(r,z)$ in streaming direction $z$  is presented for fixed $M$ and different values of the density  parameter $r_{s}$. At the highest density, $r_s=0.1$, a deep attractive minimum exists close to the ion. A density reduction (increase of $r_{s}$), reduces the number of electrons that are being deflected by the ion, which lowers the excess electron density behind the ion. Also the average number of electrons scattered with a small impact parameter (i.e. scattered closer to the ion) decreases. 
As a result, the attraction becomes weaker and the distance of the area of attraction from the ion grows.    
To make the analysis more quantitative, we plot the absolute depth of the first minimum of $\Phi$  behind the ion in Fig.~\ref{fig:MinRS}.a as a function of $M$ for different values of $r_{s}$. The corresponding location of this minimum is shown in Fig.~\ref{fig:MinRS}.b. 
\begin{figure}[h]
\vspace*{0.5cm}
\includegraphics[width=.8\linewidth]{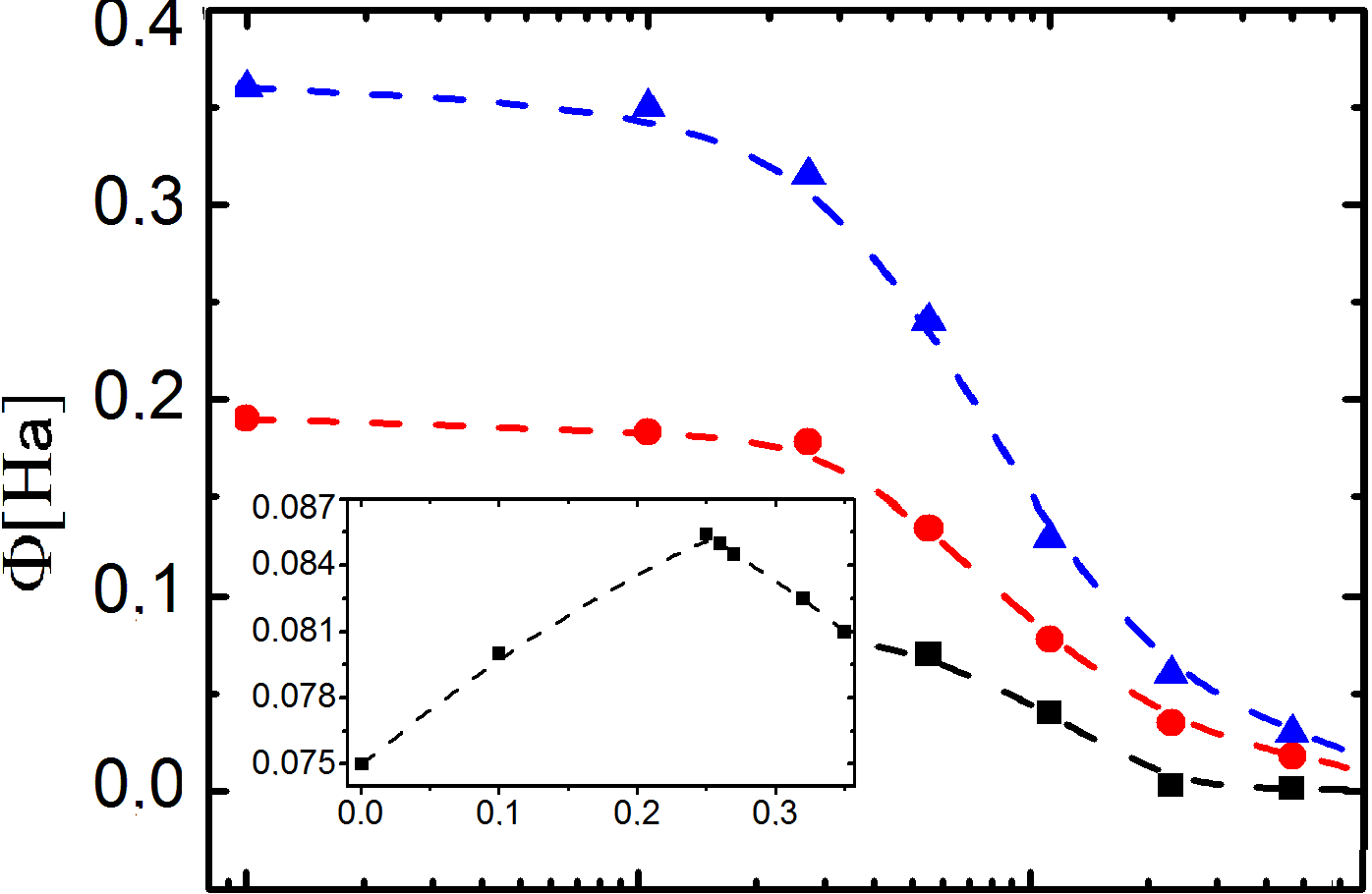}~\hspace{-0mm}\begin{scriptsize}a)\end{scriptsize}
\includegraphics[width=.79\linewidth]{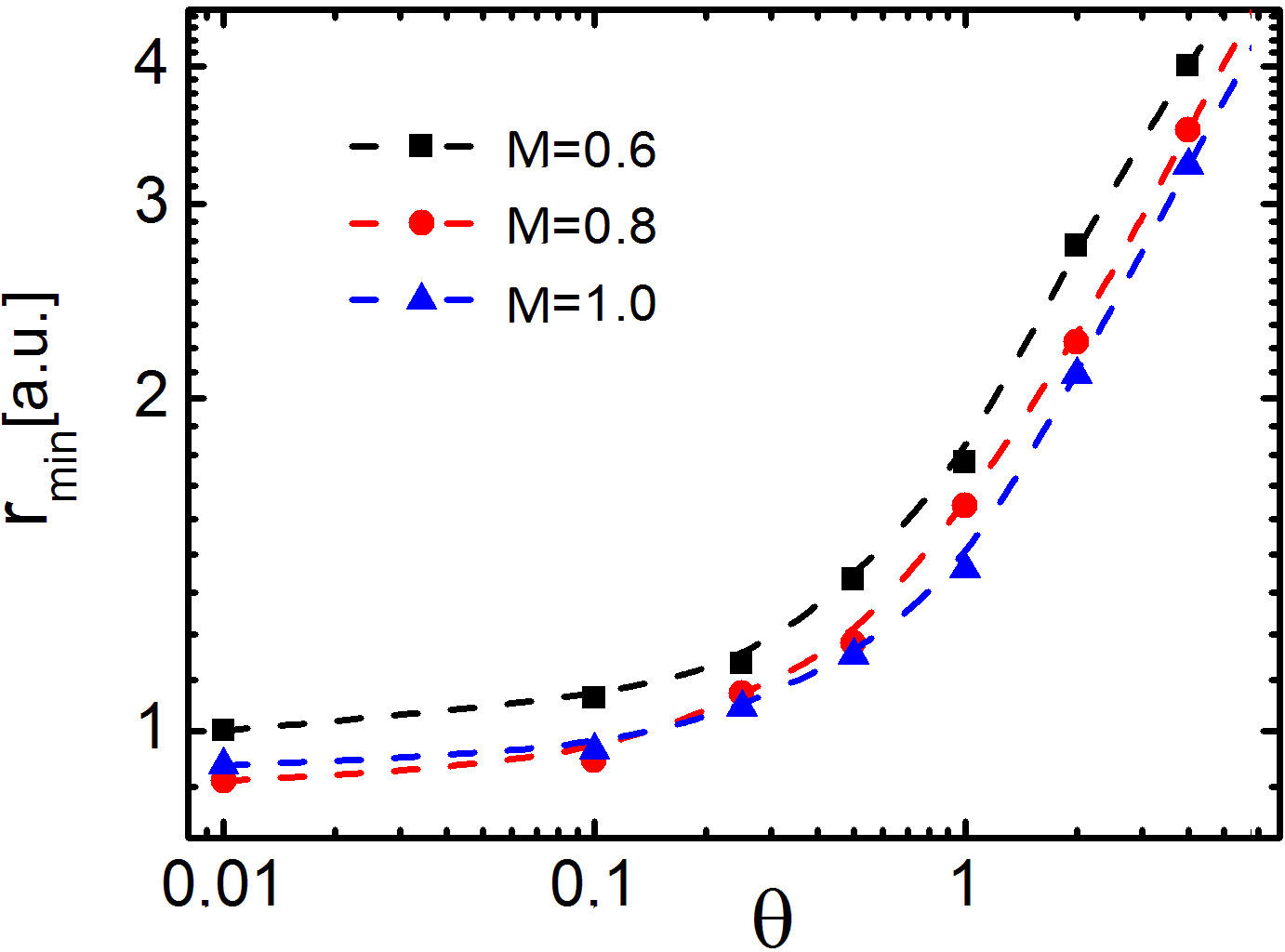}~\hspace{-0mm}\begin{scriptsize}b)\end{scriptsize}
\caption{(Color online) Absolute value (top) and location (bottom) of the first minimum of $\Phi$ behind the ion as a function of temperature for $r_{s}=0.5$ and fixed $M$ values indicated in the figure.}
\label{fig:PotVSdegpar}
\end{figure}
\begin{figure}[t]
\includegraphics[width=.945\linewidth]{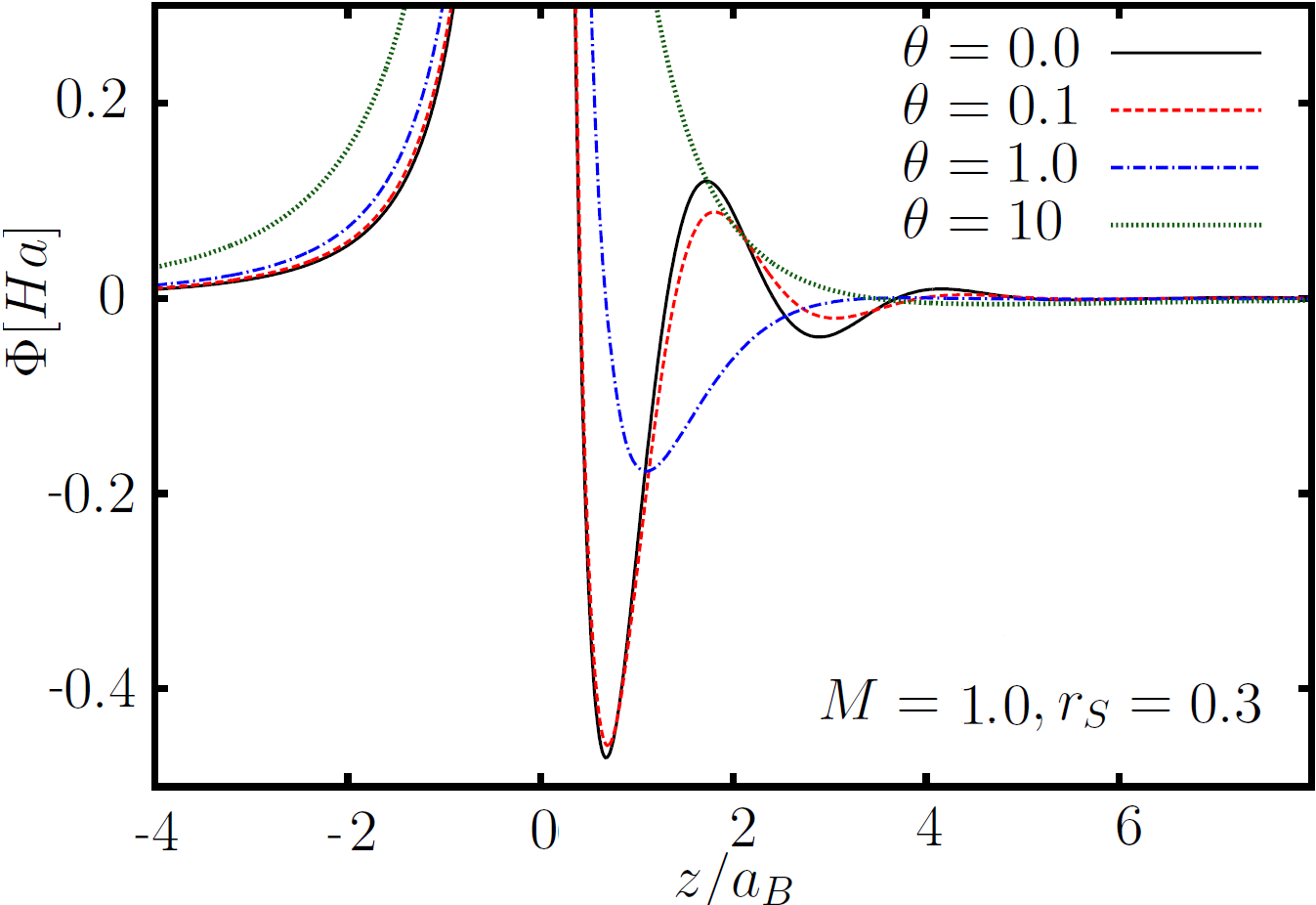}~\hspace{-0mm}\begin{scriptsize}a)\end{scriptsize}
\includegraphics[width=.95\linewidth]{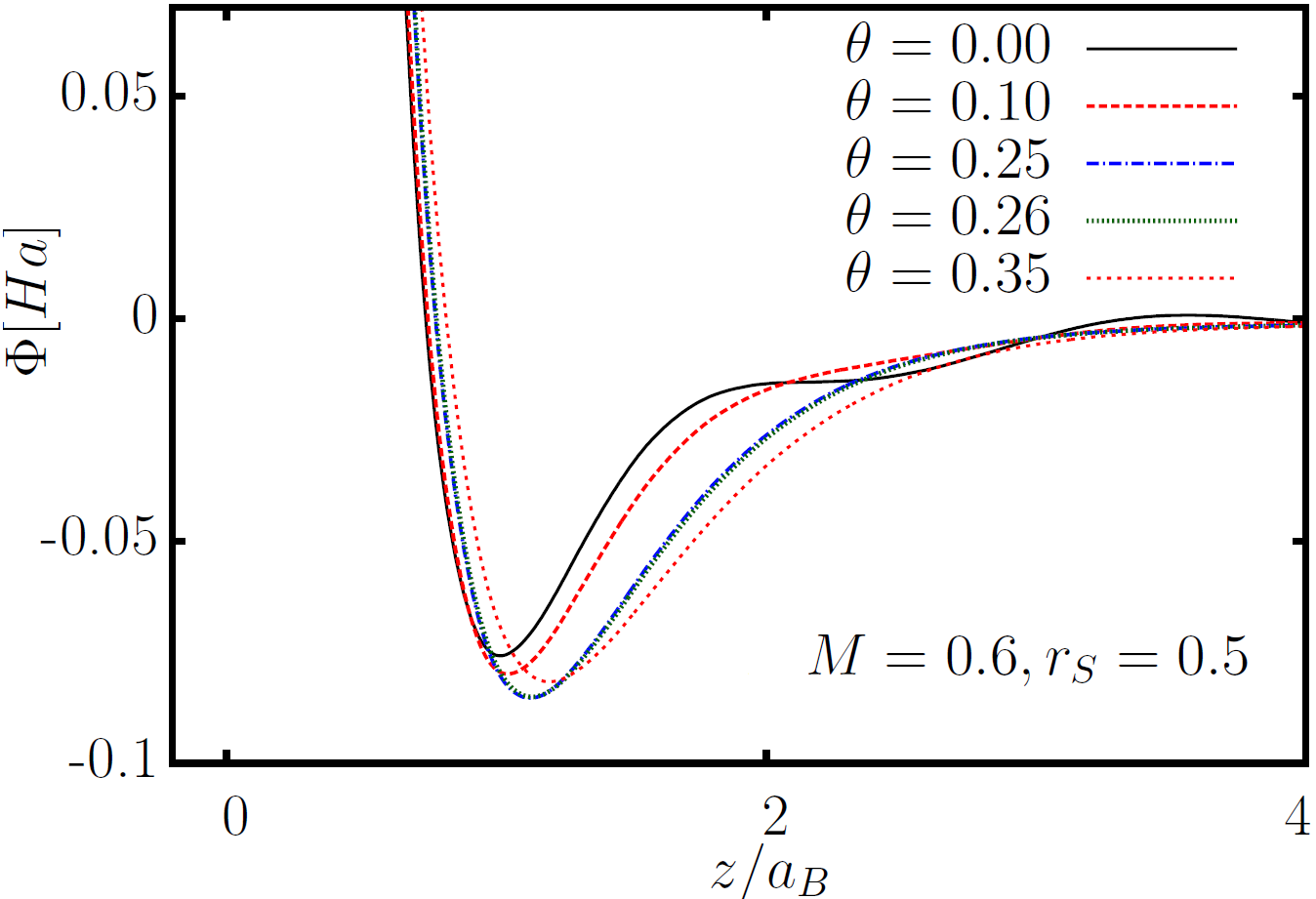}~\hspace{-0mm}\begin{scriptsize}b)\end{scriptsize}
\caption{(Color online) Temperature effect on the dynamically screened ion potential $\Phi(z)$ in streaming direction at the ion position ($r=0$). {\bf Top}: at $M=1.0$, {\bf bottom}: for at $M=0.6$. Temperatures are indicated in the figure.}
\label{fig:Mermin_M0.6_M1.0}
\end{figure}
\item[c.] {\em Effect of temperature.} The random thermal motion of electrons works against electron focusing and, thus, inhibits the creation of the wake, as is clearly seen in Fig.~\ref{fig:M}.c. 
In the limit of high temperature ($\theta \gg 1$) and relatively small streaming velocity ($M\sim 1$), the oscillatory structure is spread out by the thermal motion, and the fraction of electrons scattered in the streaming direction is strongly reduced and insufficient for the creation of a wake. In the opposite limit of low temperatures ($\theta<0.1$), the kinetic energy is dominated by the Fermi energy (see Fig.~\ref{fig:ks}.b). Changes in temperature strongly affect 
the depth of the first minimum (except at very strong degeneracy, $\Theta \lesssim 0.5$). Also, the next extrema that are clearly expressed at $\Theta=0$ are 
drastically influenced by temperature and vanish above $\Theta \approx 1$ (see Figs.~\ref{fig:M}.c and \ref{fig:Mermin_M0.6_M1.0}). Thus for the computation of the dynamically screened potential, Eq.~(\ref{POT}), correct account of finite temperature effects is crucial. 

Consider now the higher order minima of the potential. Analyzing the results of  Fig.~\ref{fig:M} and, comparing with Fig.~\ref{fig:ks}.b, it is seen that the second minimum exists only at fast streaming, for $\left< K \right>_{S} >\left< K \right>_{Q} $. 
 For $\theta<0.1$ this corresponds to $M>0.77$. Interestingly, the same condition was found for the ultra-relativistic quark-gluon plasma (QGP) \cite{Toma} where $M=v/c$ ($c$ is the speed of light, see also Appendix B.). 

Let us now come back to the main potential minimum. Its depth and location are plotted in Figs.~\ref{fig:PotVSdegpar}.a and~\ref{fig:PotVSdegpar}.b, respectively. One clearly sees a monotonic reduction (increase) of the potential depth (distance from the ion) with increasing temperature,  in the range $\theta \gtrsim 0.1$. For lower temperatures the depth and minimum position saturate. 
An unexpected observation is that the depth may, for certain parameter combinations, {\em increase with temperature}. 
Figures~\ref{fig:PotVSdegpar} and \ref{fig:Mermin_M0.6_M1.0}.b. show that the potential depth grows monotonically with temperature from $\theta=0$ to $0.25$, after which the depth decreases again. A simple explanation of this non-monotonic dependence is that, with increasing temperature, particles
become more classical (poin-tlike) which increases the scattering effect of electrons by
the ion \cite{q_crystal_melting}.
\end{description}

\subsection{Analysis of collision effects on the screened potential}\label{ss:collisions}
\begin{figure}[t]
\vspace*{0.5cm}
\includegraphics[width=0.95\linewidth]{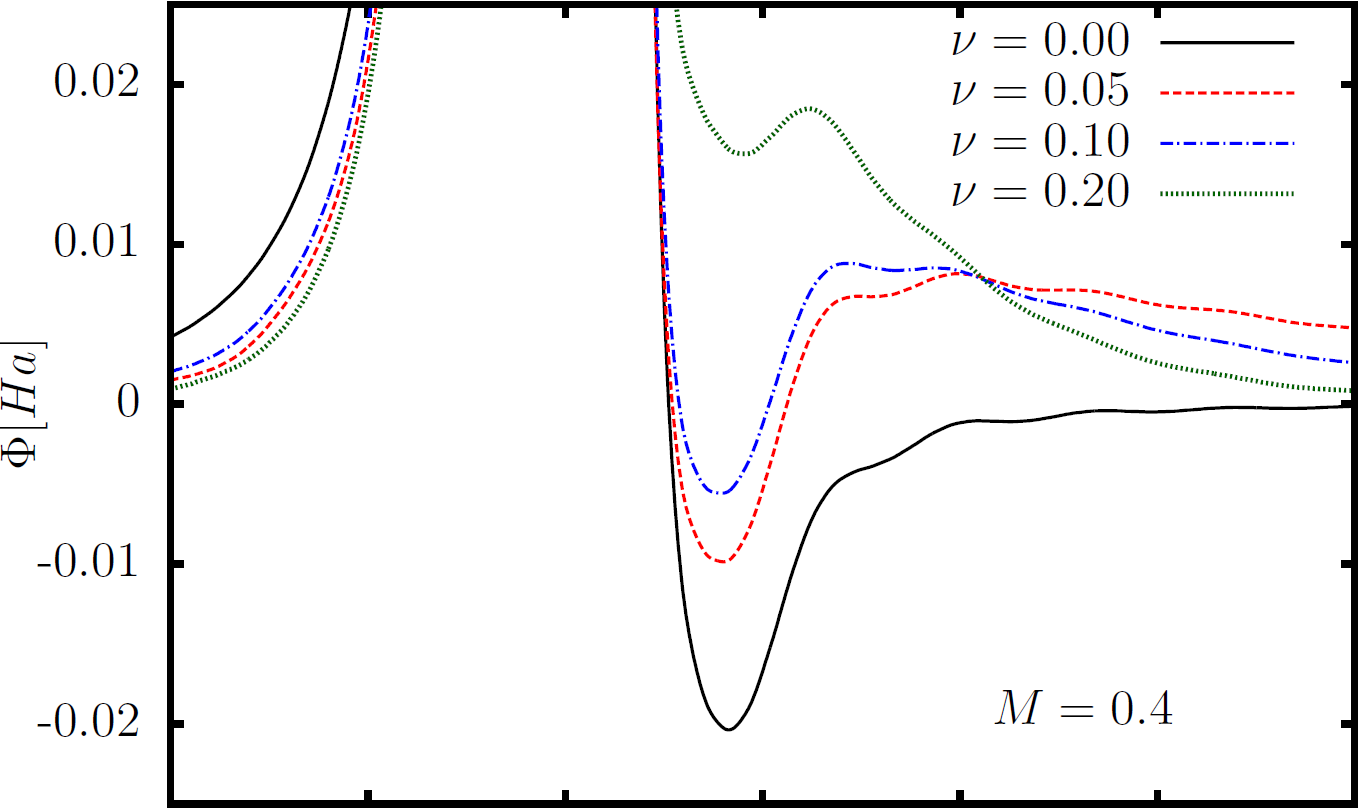}~\hspace{-0mm}\begin{scriptsize}a)\end{scriptsize}
\includegraphics[width=0.95\linewidth]{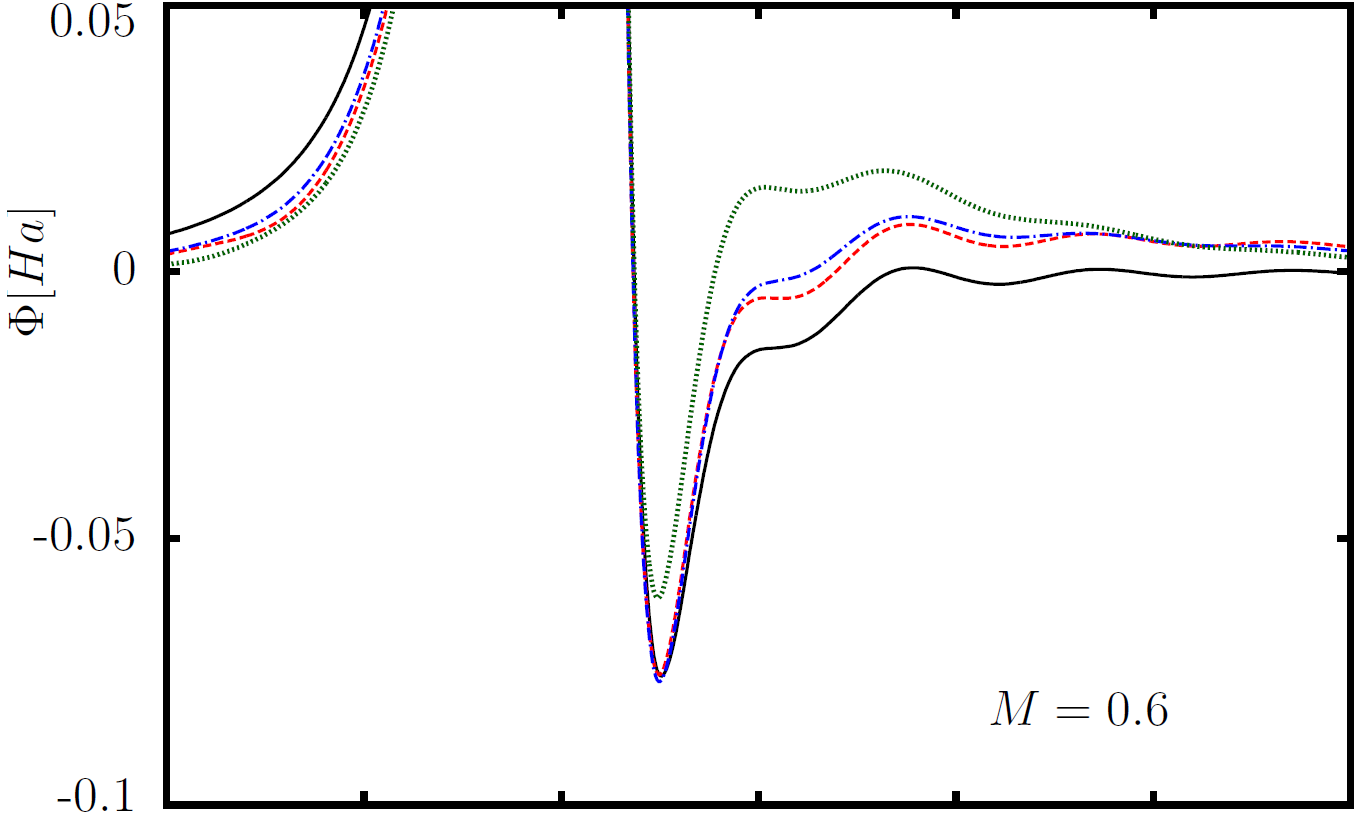}~\hspace{-0mm}\begin{scriptsize}b)\end{scriptsize}
\includegraphics[width=0.95\linewidth]{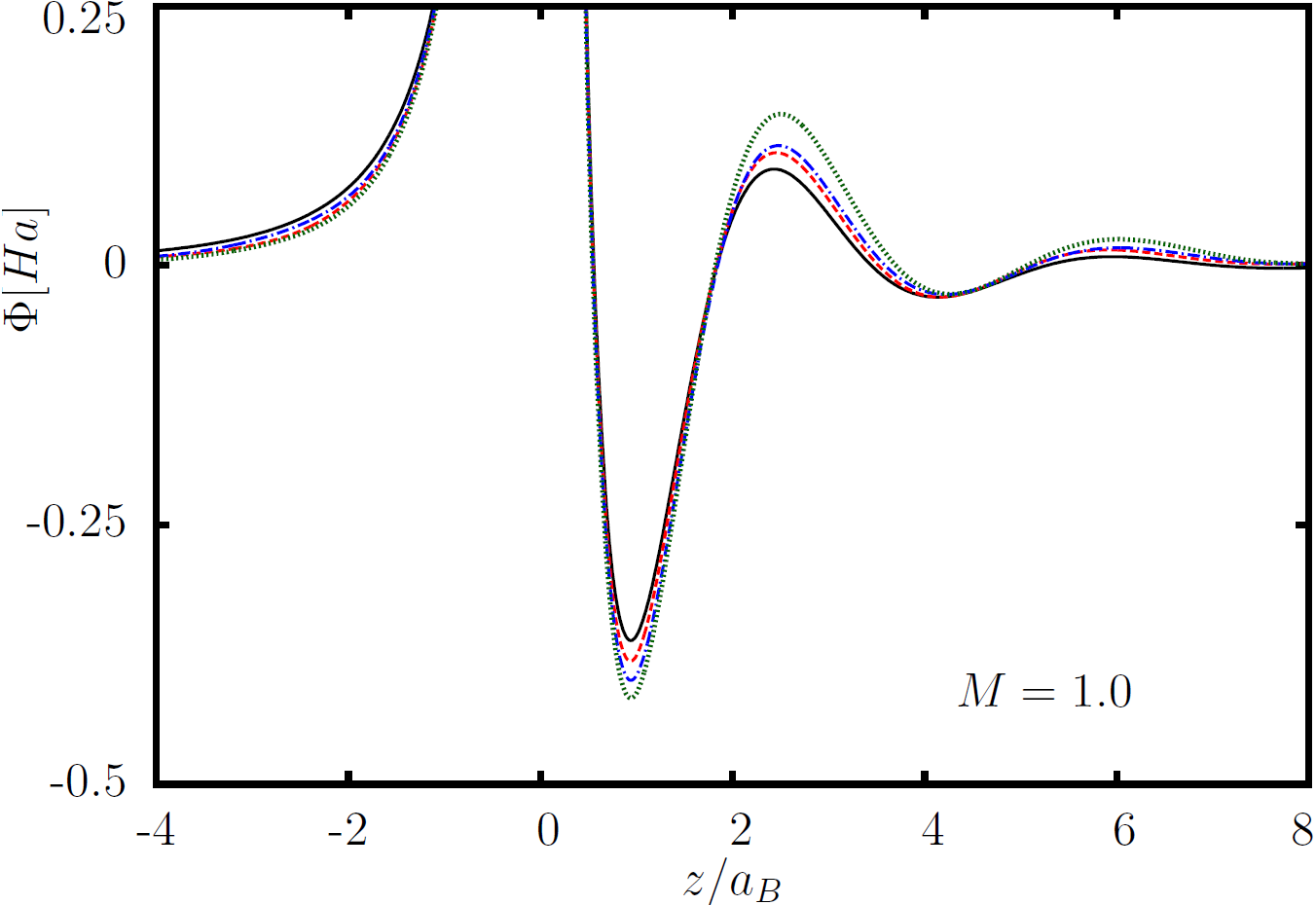}~\hspace{-0mm}\begin{scriptsize}c)\end{scriptsize}
\caption{(Color online) Effect of collisions on the dynamically screened ion potential $\Phi(z)$ for
{\bf Top}: $M=0.4$, {\bf center}: $M=0.6$, {\bf bottom}: $M=1.0$. The potential is shown in streaming direction at the ion position ($r=0$).} 
\label{fig:Mcoll1}
\end{figure}
\begin{figure}[t]
\vspace*{0.5cm}
\includegraphics[width=0.45\textwidth]{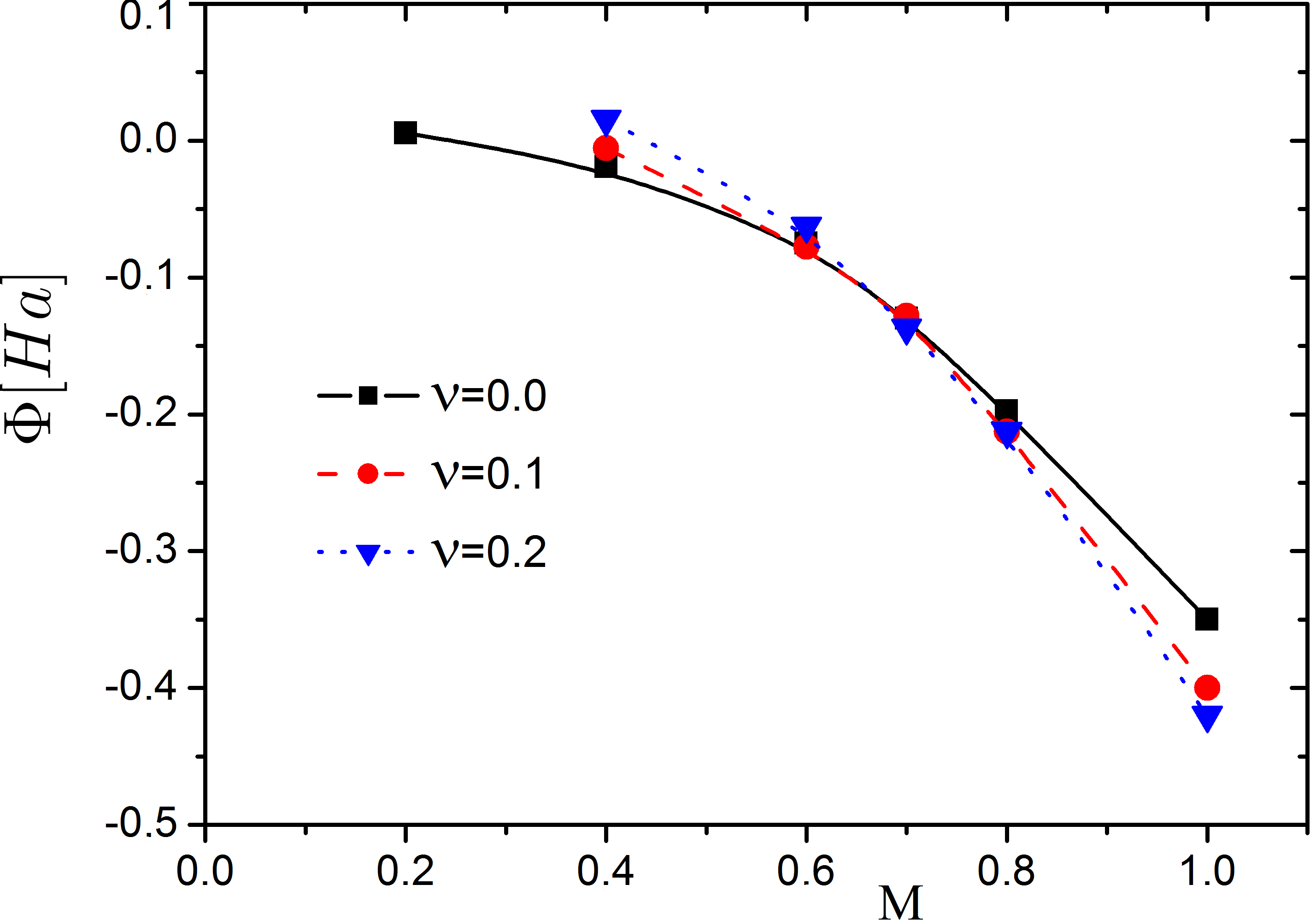}
\caption{(Color online) Effect of collisions on the depth of the first minimum of $\Phi$ behind the ion, as a function of $M$ for $r_{s}=0.5$. Note the non-monotonic effect of collisions leading either to a reduction ($M>0.65$) or an enhancement ($M<0.65$) of the potential minimum compared to the collisionless case (full black curve).}
\label{fig:Potvscoll}
\end{figure}
\begin{figure}[t]
\includegraphics[width=1\linewidth]{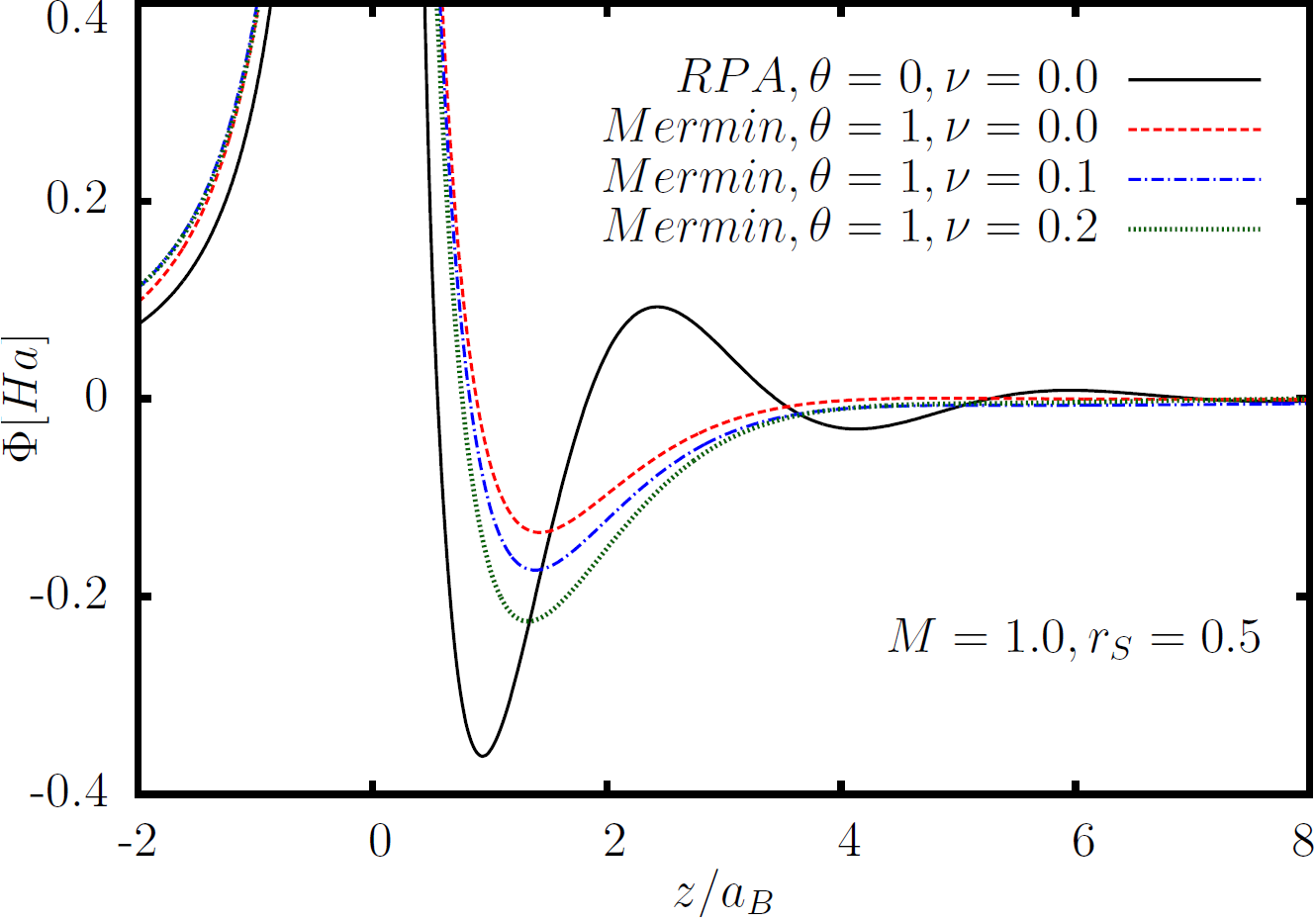}
\caption{(Color online) Combined effect of finite temperature and collisions on the dynamically screened ion potential $\Phi(z)$. The result of the Mermin dielectric function for a plasma at $\Theta=1$ and $r_s=0.5$ is shown for the example $M=1$ and compared to the  
result using the RPA dielectric function in the zero-temperature limit (full black line).}
\label{fig:MerminVsRPA}
\end{figure}
Until now all results for the screened potential were computed using the finite-temperature Mermin dielectric function with collisions included selfconsistently via formula (\ref{nu0}). As we discussed in Section \ref{s:mermin} this formula most likely underestimates the electron collision frequency $\nu$ in warm dense matter.
We, therefore, separately analyze the effect of the collision frequency on our results, allowing for a broader range of values for $\nu$. 

In Fig.~\ref{fig:Mcoll1} the effective dynamically screened ion potential is shown for different collision frequencies at $M=0.4$, $M=0.6$, and $M=1.0$.
Obviously, inclusion of collisions changes the shape of the potential dramatically. Depending on the streaming parameter collisions may reduce the depth of the first minimum ($M<0.65$) or even enhance the minimum ($M>0.65$). The biggest effect of collisions is observed for slowly streaming particles (small $M$-values) since there the kinetic energy of the directed motion is low and the particle trajectories are more strongly affected by collisions.

The influence of collisions on the depth of the first minimum of the dynamically screened potential is summarized in Fig.~\ref{fig:Potvscoll}. In contrast to the depth, the position of the first minimum does practically not change with the collision frequency.

Finally, in Fig.~\ref{fig:MerminVsRPA} we summarize the influence of collisions and finite temperature on the dynamically screened potential. For comparison we also include the zero-temperature RPA dielectric function which has been used in most earlier works devoted to wake effects in quantum degenerate systems. Obviously the latter result is very inaccurate, largely overestimating the effective ion-ion attraction. Both, finite temperature and collisions lead to a reduction of the wake effects. This clearly confirms the importance of using the correct dielectric functions for streaming plasmas in the warm dense matter regime.

\subsection{Screened potential away from the symmetry axis}\label{ss:off-axis}
After anlyzing the ion potential in streaming direction at the ion location let us now also consider its shape in perpendicular direction. This is shown in Figs.~\ref{fig:wake}, \ref{fig:wake_rs} and \ref{fig:wake_M}. The anisotropy of this potential is striking, in marked difference to the static Yukawa potential (\ref{POT_stat}).

We first consider the transition from the classical limit, $k_{B}T \gg E_{F}$, to the quantum limit, $E_{F} \lesssim k_{B}T$, at constant density ($r_s=0.5$) and 
streaming velocity ($M=1$). The result is shown in Fig.~\ref{fig:wake}. At high temperature, $\theta= 10$ (left figure) the potential has no attractive part and only a small anisotropy indicates the existence of streaming electrons. When the temperature is reduced by a factor $10$ ($\theta=1$, middle Figure), the potential becomes strongly anisotropic and develops an attractive region behind the ion, as discussed before. The most striking observation is that the attractive area exists not only behind the ion on the $z$-axis, but also in a broad region in perpendicular direction, away from the axis. Such a structure is found to be stable in a  plasma with degenerate electrons ($r_{s}<1, \;\theta \lesssim 0.1$). Upon further reduction of temperature to $\theta=0.1$ (right figure)
the angular spread of the attractive minimum increases further and persists also for the second (repulsive) and third (attractive) extremum.

One particularly interesting observation in Fig.~\ref{fig:wake} is the striking difference of the shape of the wake pattern, as compared to the one known from classical plasmas as well as to the wake behind an object in a 
streaming fluid. While the latter cases exhibit a characteristic ``V''-shape potential pattern that opens in flow direction, in the present case of a dense quantum plasma the wake has a qualitatively different shape: the wakes bend {\em towards the ion}. This appears to be a pure quantum effect that is related to the finite extension of the electron wave function combined with its anisotropy in the case of streaming. This explanation is supported by very similar behavior of the ultra-relativistic quark gluon plasma reported by Thoma {\em et al.} in Refs.~\cite{Toma,Toma2}. They performed a zero-temperature RPA calculation using a color-Coulomb potential. The main results are discussed in Appendix B, for a discussion of the underlying physics or the QGP, see Refs.~\cite{filinov_cpp12, filinov_cpp14}. 
%

We now study the evolution of the potential shape in the strongly degenerate case ($\theta=0.1$) upon variation of the density. In Fig.~\ref{fig:wake_rs} we show, for fixed streaming velocity $M=1$, three cases corresponding to $r_s=0.1$ (left), $r_s=0.3$ (middle) and $r_s=1$ (right), spanning three orders of magnitude in density. Interestingly, in all three cases the potential pattern looks very similar, the main difference being a scaling of the absolute length scales: with increasing density 
the pattern is compressed in all directions thereby retaining the characteristic bend toward the ion. 
\begin{widetext}

The formation of this quantum wake pattern is particularly obvious when, in the quantum high-density regime, the streaming velocity is increased. In Fig.~\ref{fig:wake_M}
we show calculation results for fixed values of density and temperature, $r_s=0.3, \; \theta=0.1$ and three streaming velocities: $M=0.4$ (left figure part), $M=0.6$ (middle) and $M=1$ (right). With increasing flow velocity the perpendicular spread of the attractive region grows continuously and increases its curvature towards the ion. Interestingly, the bending towards the ion is already visible for $M=0.4$ just when the attractive minimum on the $z$-axis emerges, cf. Fig.~\ref{fig:M}.a. indicating that this is a generic property of quantum plasmas.

\begin{figure}[h]
\includegraphics[width=0.302\textwidth]{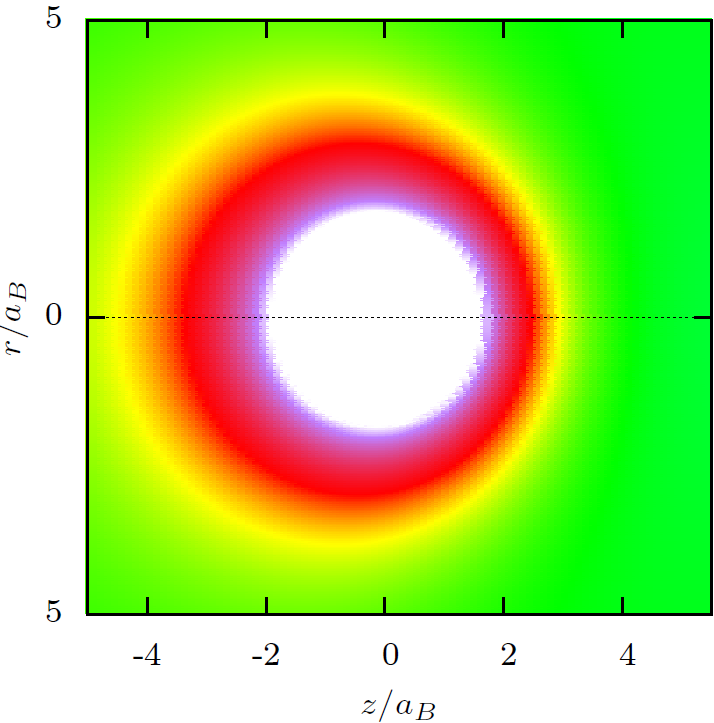}~\hspace{-4.7mm}
a)
\includegraphics[width=0.2685\textwidth]{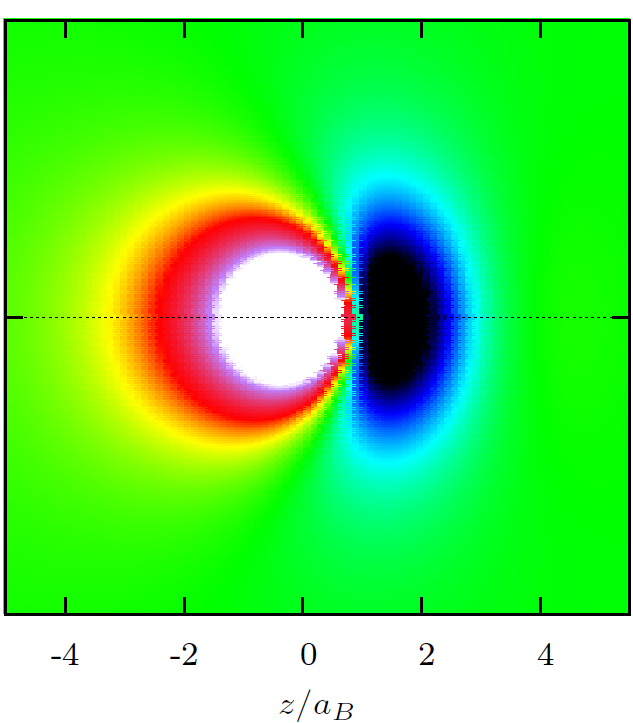}~\hspace{-4.3mm}
b)
\includegraphics[width=0.3345\textwidth]{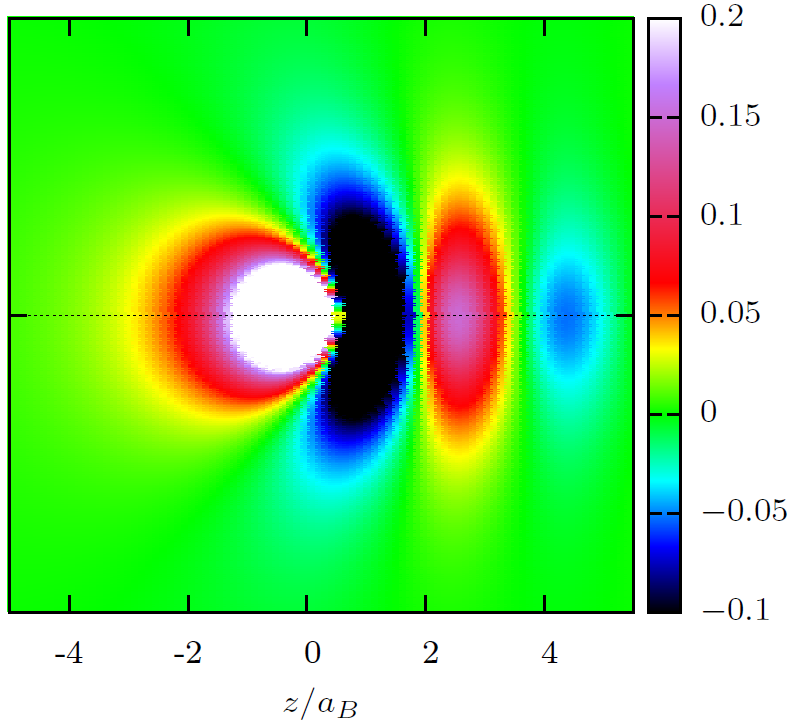}~\hspace{-18mm}
c)
\caption{(Color online) Dynamically screened ion potential $\Phi(r,z)$ for $r_{s}=0.5$ and $M=1$ and three different temperatures:
 a) $T=2.32\times 10^7 K (\theta=10)$, b) $T=2.32\times 10^6 K  (\theta=1)$, and c) $T=2.32\times 10^5 K (\theta=0.1)$. 
The ion is located at $\{r,z\}=\{0,0\}$. The electrons stream in $z$-direction from left to right. }
\label{fig:wake}
\end{figure}

\begin{figure}[t]
\includegraphics[width=0.304\textwidth]{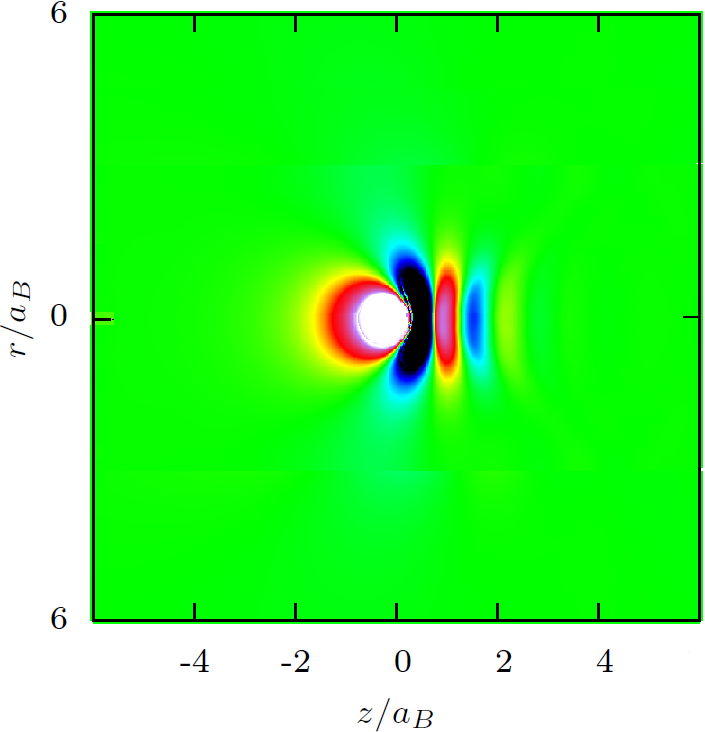}~\hspace{-4.7mm}
a)
\includegraphics[width=0.265\textwidth]{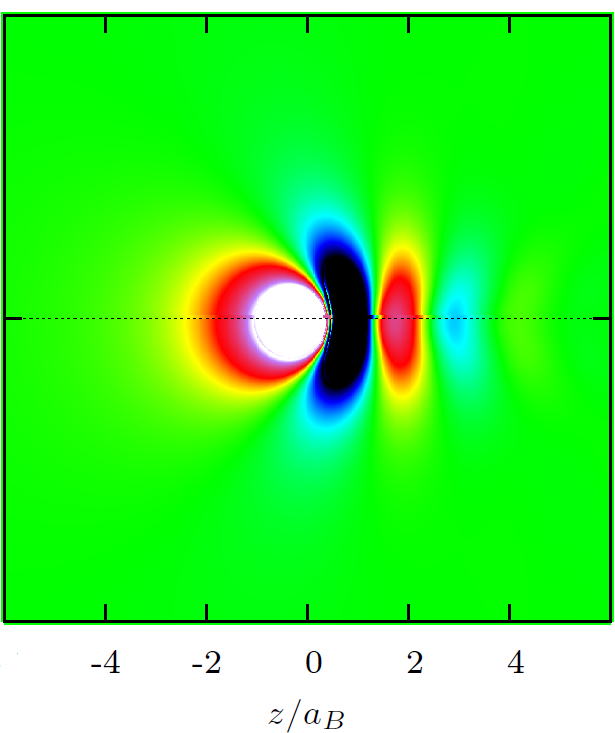}~\hspace{-4.3mm}
b)
\includegraphics[width=0.33\textwidth]{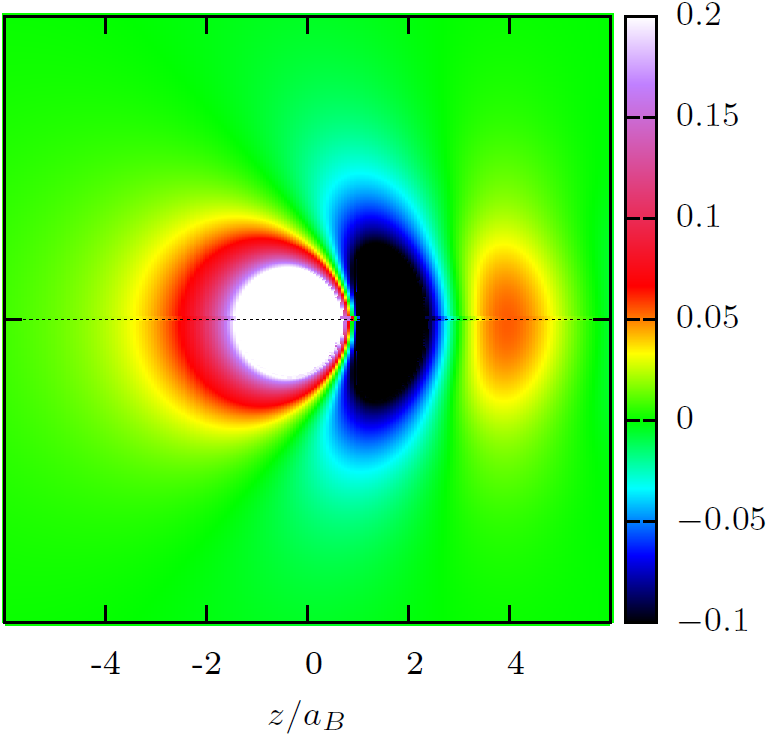}~\hspace{-18mm}
c)
\caption{(Color online) Dynamically screened ion potential $\Phi(r,z)$ for $\theta=0.1$ and $M=1$ and three different densities:
 a) $r_s= 0.1$, b) $r_s=0.3$, and c) $r_s=1.0 $. 
The ion is located at $\{r,z\}=\{0,0\}$. The electrons stream in $z$-direction from left to right.}
\label{fig:wake_rs}
\end{figure}



\subsection{Perturbed electron density}\label{ss:e-density}
Having computed the dynamically screened ion potential we can now directly obtain the local electron density, as explained in Sec.~\ref{s:mermin}. 
We expect that the dynamically screened ion potential will selfconsistently perturb the electron density, according to Eq.~(\ref{eq:e-density}). In particular, regions of an attractive potential should give rise to a local enhancement of the electron density. The results of evaluation of Eq.~(\ref{eq:e-density}) and a subsequent three-dimensional Fourier transformation are shown in Figs.~\ref{fig:density_focus} and \ref{fig:density_focus_rs}. In Fig.~\ref{fig:density_focus} we show results for the high-density quantum case with $\theta=0.1$ and $r_s=0.3$ and three velocities: $M=0.4$ (left figure part), $M=0.6$ (middle) and $M=1$ (right). At zero streaming velocity the electron distribution is, of course, isotropic around the ion. For $M=0.4$, remnants of an isotropic distribution are still visible upstream (to the left) of the ion and in perpendicular direction, but behind the ion already a small density enhancement is visible. This pattern is very close to the one of the attractive region of the potential, cf. Fig.~\ref{fig:wake_M}.a, and is also bent towards the ion. This trend continues for $M=0.6$ where an area of excess density is formed. Finally, for $M=1$ a very strong and broad density peak is formed that resembles the potential but is more elongated in flow direction. Also, a second spot of reduced electron density has formed, consistent with the second repulsive maximum of the ion potential in Fig.~\ref{fig:wake_M}.c.

Next, in Fig.~\ref{fig:density_focus_rs} we consider the density perturbation at fixed temperature and streaming velocity, for three values of the electron density. The behavior is again similar to the effective potential that was shown for the same parameters in Fig.~\ref{fig:wake_rs}. With a reduction of the density the length scales of the electron density pattern increase.
Thus, our simulations clearly confirm our previous discussion about the focusing effect experienced by the electrons from the ion. The magnitude of the effect is substantial at high flow velocities. According to the color scale in Fig.~\ref{fig:density_focus_rs}.c, at $M=1$, the peak height of the density perturbation  
is $|\delta n|/n_0 \lesssim 0.05$ and even smaller in all other cases.
This again confirms that the linear response condition is well fulfilled.


%
%

\begin{figure}[t]
\includegraphics[width=0.3\textwidth]{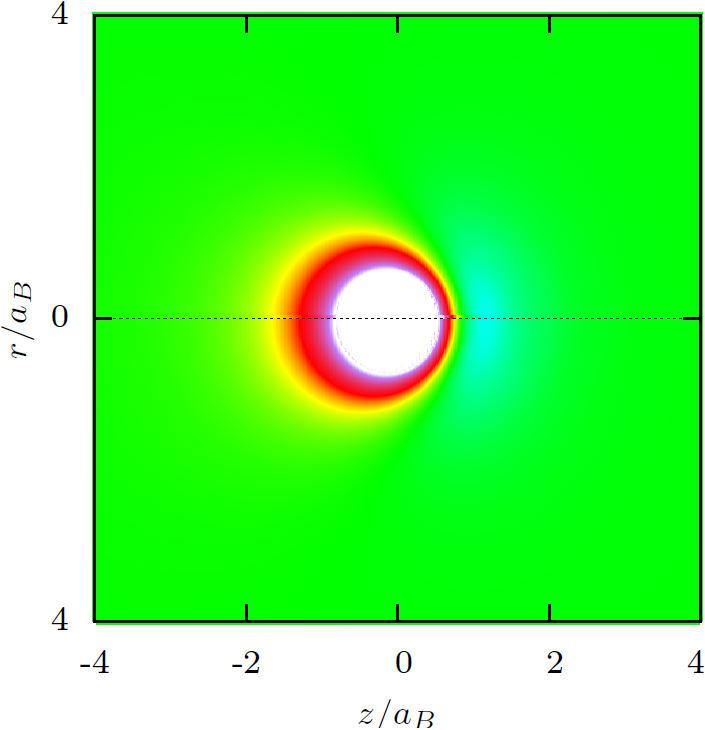}~\hspace{-4.7mm}
a)
\includegraphics[width=0.26\textwidth]{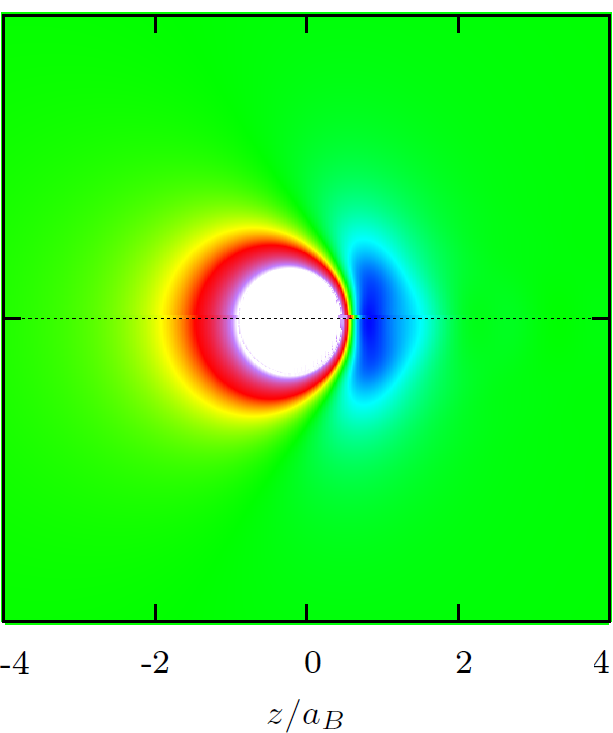}~\hspace{-5.7mm}
b)
\includegraphics[width=0.332\textwidth]{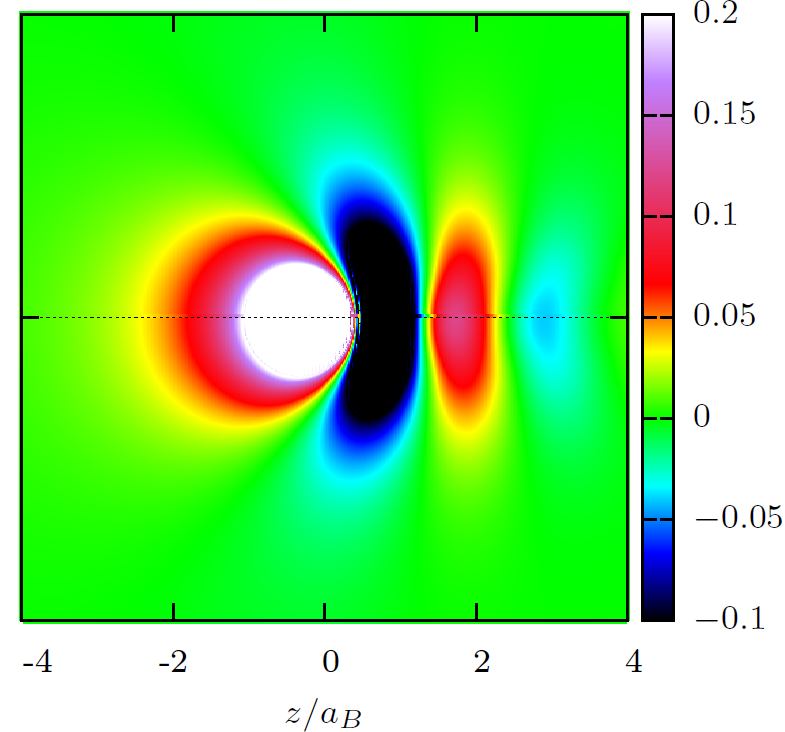}~\hspace{-18mm}
c)
\caption{(Color online) Dynamically screened ion potential $\Phi(r,z)$ for $\theta=0.1$ and $r_s=0.3$ for three different streaming velocities:  a) $M=0.4 $, b) $M=0.6 $, and c) $M=1.0 $. 
The ion is located at $\{r,z\}=\{0,0\}$. The electrons stream in $z$-direction from left to right.}
\label{fig:wake_M}
\end{figure}

\begin{figure}[t]
\includegraphics[width=0.29\textwidth]{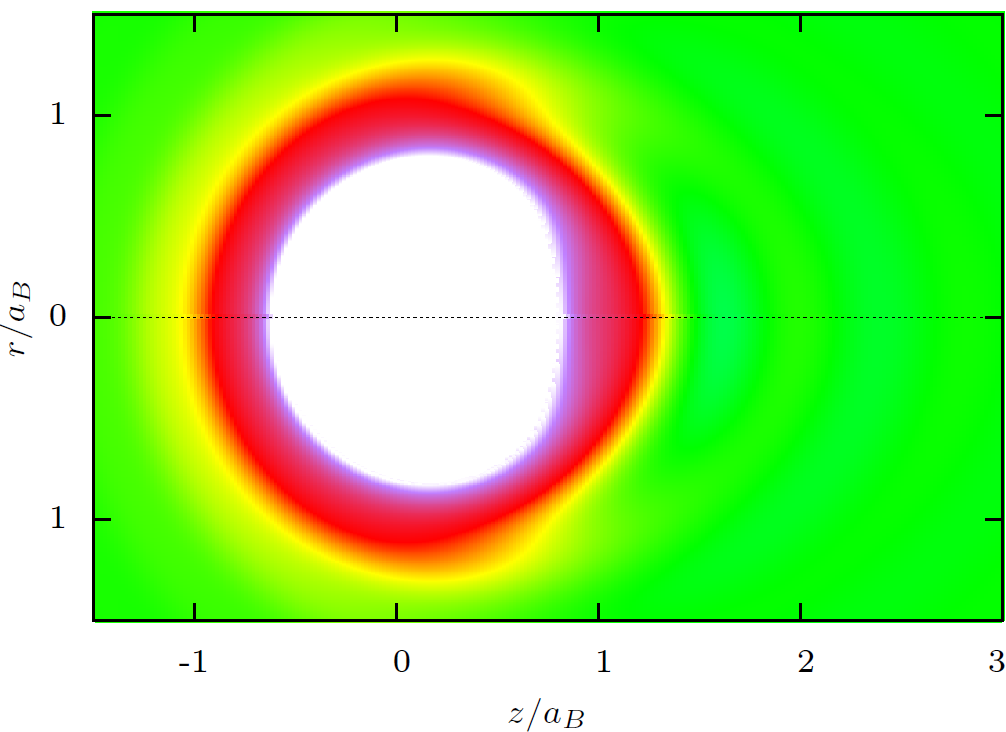}~\hspace{-4.7mm}
a)
\includegraphics[width=0.265\textwidth]{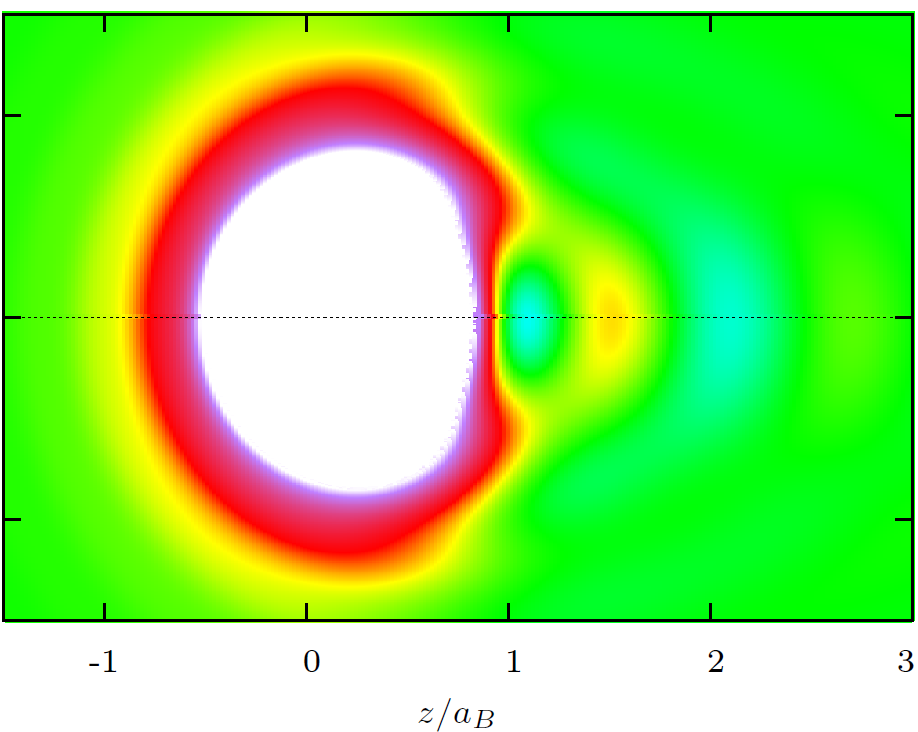}~\hspace{-4.3mm}
b)
\includegraphics[width=0.345\textwidth]{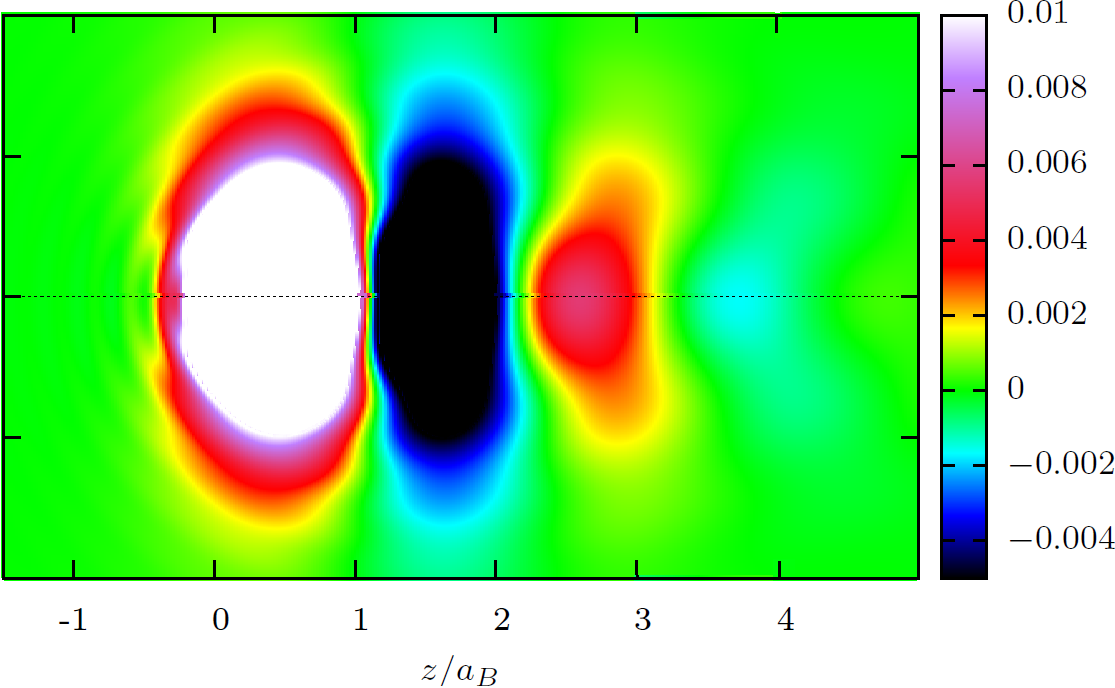}~\hspace{-18mm}
c)
\caption{(Color online) Electron density perturbation (in units of the unperturbed density $n_0$) produced by the focusing effect of the ion for $\theta=0.1$ and $r_s=0.3$ for three different streaming velocities:  a) $M=0.4 $, b) $M=0.6 $, and c) $M=1.0 $. Same parameters as in Fig.~\ref{fig:wake_M}.}
\label{fig:density_focus}
\end{figure}

\begin{figure}[t]
\includegraphics[width=0.30\textwidth]{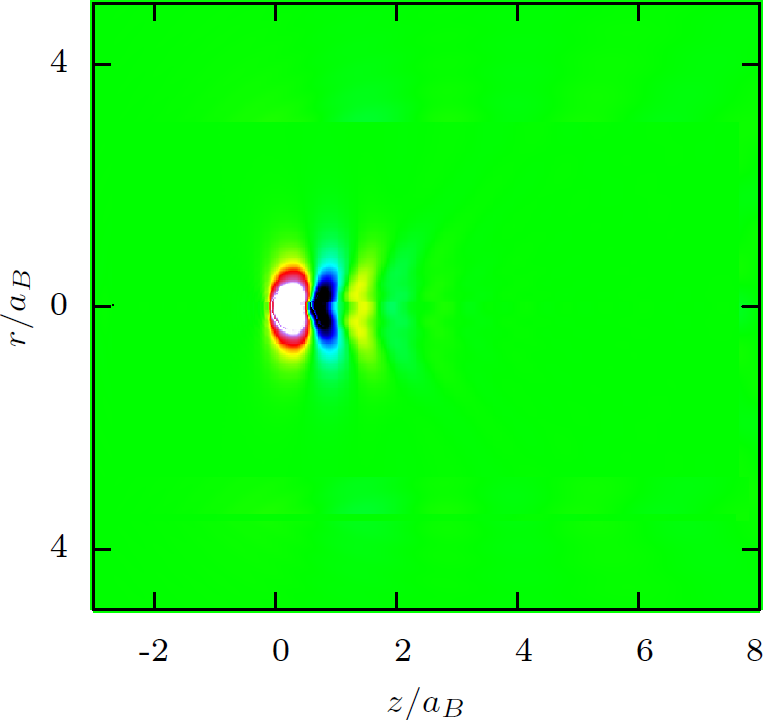}~\hspace{-4.7mm}
a)
\includegraphics[width=0.265\textwidth]{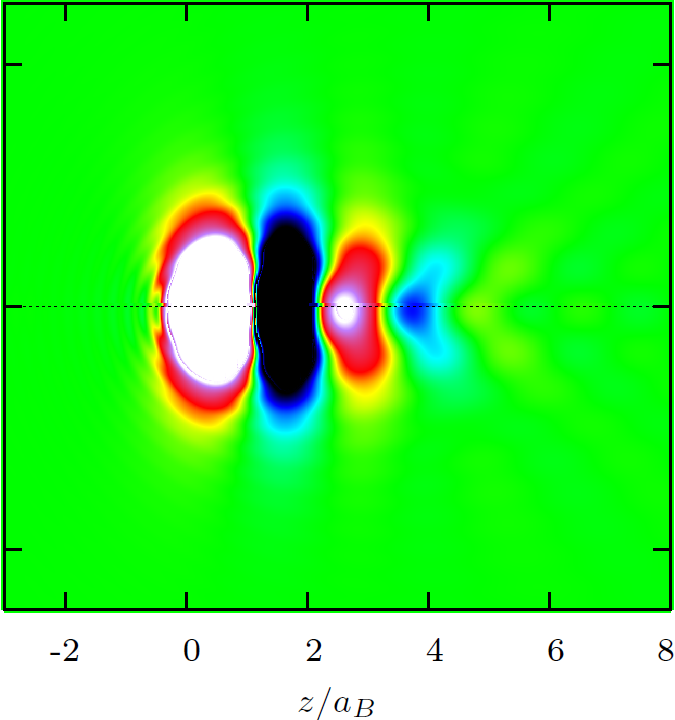}~\hspace{-4.3mm}
b)
\includegraphics[width=0.335\textwidth]{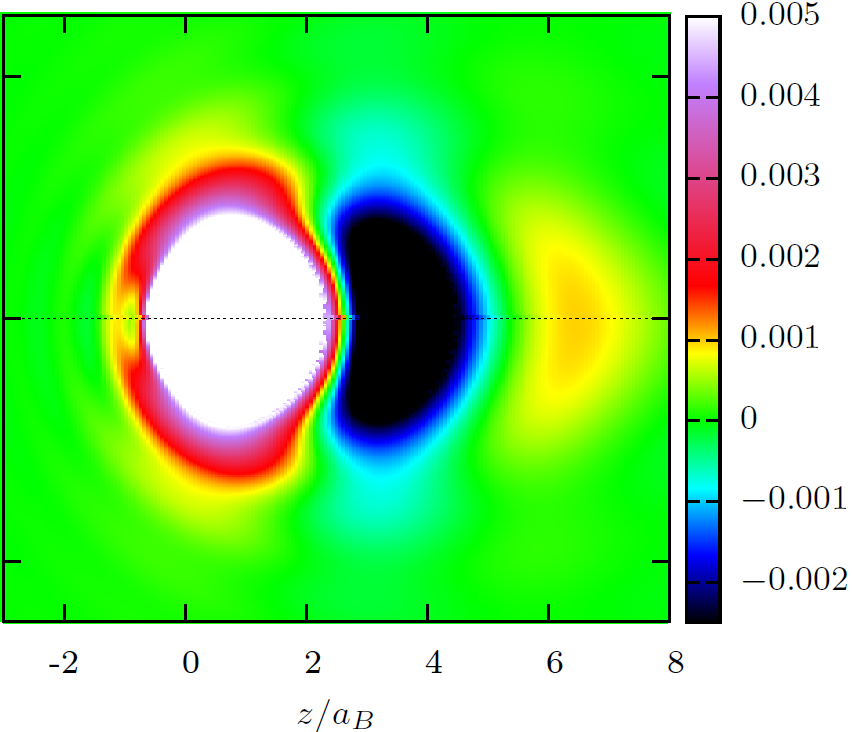}~\hspace{-18mm}
c)
\caption{(Color online) Electron density perturbation (in units of the unperturbed density $n_0$) produced by the focusing effect of the ion for $\theta=0.1$ and $M=1$ for three different electron densities:  a) $r_s=0.1 $, b) $r_s=0.3 $, and c) $r_s=1.0 $. For better comparison, in Fig.c the density is reduced by a factor $10$. Same parameters as in Fig.~\ref{fig:wake_rs}.}
\label{fig:density_focus_rs}
\end{figure}

\end{widetext}

\section{Summary and Outlook} \label{s:dis}
This paper is devoted to dense electron-ion plasmas in the warm dense matter regime where electrons (ions) are weakly (strongly) correlated and electronic quantum effects are relevant. We concentrated on stationary nonequilibrium states where electrons move relative to the ions---a situation that is ubiquitous in dense plasmas including electron or ion beams, laser accelerated electrons or ions penetrating a dense quantum plasma or a metal (ion stopping). 

Using a multiscale approach \cite{Patrick2} the problem was reduced to the computation of a dynamically screened ion potential that, in a next step, can be used directly in classical molecular dynamics simulations of the ions. This scheme was successfully applied to classical dusty plasmas before \cite{Patrick1,Patrick4} and is here extended to dense quantum plasmas. In this paper we concentrated on the first step of the approach, presenting 
the dynamically screened ion potential and the electron density perturbation in a dense quantum plasma in the presence of streaming degenerate electrons. We used the  Mermin dielectric function, achieving a two-fold improvement of similar previous studies: we included both, finite temperature effects and collisions. 

We demonstrated that the zero-temperature approximation is not adequate for situations where the temperature exceeds about 20 percent of the Fermi energy, cf. Figs.~\ref{fig:PotVSdegpar} and \ref{fig:Mermin_M0.6_M1.0}. Equally important is the correct account of collisions: the collisionless approximation (RPA) drastially overestimates wake effects, and collisions are particularly important at moderate streaming velocities, $M \lesssim 0.7$, cf. Figs.~\ref{fig:Mcoll1} and \ref{fig:Potvscoll}. The precise value of the collision frequency $\nu$ depends on the type of quantum plasma, excitation conditions and relevant collision processes. Thus the results obtained in this paper depend not only on the dimensionless parameters $r_s, \Theta, M$, but also on $\nu$.

We presented a detailed analysis of the ion potential that exhibits wake effects giving rise to an effective ion-ion attraction at short distances.
This effect is similar to wakes behind an object in a moving fluid, wakes in laser plasmas or in dusty plasmas which all are able to accelerate particles -- in this case, a second ion. The physical mechanisms are similar in all cases -- it is the attraction between a heavy test particle and the streaming light particles giving rise to a deflection of the latter and, eventually, to an excess density behind the test particle. This interpretation was directly confirmed by computing the electron density perturbation which reveals a clear enhancement behind the ion.

Our results revealed a substantial strength of the attractive potential. The depth of the first (main) potential minimum reaches values of more than one Hartree, cf. Fig.~\ref{fig:MinRS}.a, which may have a profound effect on the structure of strongly correlated ions in dense low-temperature plasmas. At the same time the minimum is separated from the ion significantly more than the mean interparticle distance, cf. Fig.~\ref{fig:MinRS}.b. This will prevent a large scale ordering of ions at such distances as claimed in Ref. ~\cite{shukla_prl12}, see also Ref.~\cite{bonitz_pre13}. Only at small length scales and/or during nonequilibrium processes 
where charge neutrality is violated and the electron density exceeds the ion density this attraction may, eventually, play a  role in the plasma dynamics. But this remains a subject of further investigation.

Aside from the similarities of the wakes in quantum plasmas to those in classical systems, there are also qualitative differences. The most striking one is the different shape of the wake pattern behind the ion which does not have a ``V''-shape but is bent upstream. This is a pure quantum diffraction effect that is related to the finite extension of the electron wave function which is of the order of the thermal DeBroglie wavelength. 

As a side remark we note that wake effects in streaming quantum plasmas have also been found 
in quantum hydrodynamic theory (QHD). However, there the wake pattern has a characteristic ``V''-shape, in striking contrast to our results. This indicates that kinetic effects are crucial for a reliable description of streaming quantum plasmas. Furthermore, recent QHD results predicted an effective attractive ion potential, even for the case $u_e=0$ \cite{shukla_prl12}. This is again in contrast to our kinetic approach where, in the limit $u_e\to 0$, the wake effects vanish and an isotropic Yukawa-type potential is recovered. Thus our results clearly confirm the recent explanations given in Refs.~\cite{bonitz_pre13,bonitz_psc13} of the invalidity of linearized QHD for the problem of ions in streaming quantum plasmas.

Another interesting and counter-intuitive result that was not reported before is that, in the range of low temperatures, the main attractive minimum of the dynamically screened potential may become deeper with increasing temperature. Also, we observed that the account of collisions does not always simply cause damping of the potential oscillations but may, in some cases, even increase the potential depth. Both effects are due to the finite extension of the electron wave function that reduces with temperature or thermal fluctuations.

Let us now discuss the validity of our results. Our approach is based on a consistent and conserving dielectric function, so it is expected to be reliable within the validity limits of linear response theory (see below). 
Even though our approach includes some correlation effects, its validity in the range of strong correlations, $r_s>1$, is questionable. We, therefore, concentrated on the high-density case of non-relativistic electrons with $0.1 \lesssim r_s \lesssim 1$. An important consistency check is provided by comparison of the depth of the first minimum of the potential with the kinetic energy of the electrons. As can be seen in Table 1 the potential energy related to the first minimum in $\Phi(\vec r)$ is always smaller than the quantum kinetic energy of the electrons, for $r_{s}<1$. This means that the use of the linear response approximation for the present dense plasma parameters is justified. 
%
%

As an outlook we note that the importance of wake effects in streaming quantum plasmas is expected to persist also at lower density, $r_s>1$. However, here the theoretical description is essentially more complicated. Lower density reduces the quantum kinetic energy of the electron and thus increases the probability of electron capture by the ion. In Ref.~\cite{Else2} bound states near a moving ion in a fully degenerate electron plasma were considered, and it was shown that, in order to create at least one static bound state, the density parameter $r_{s}$ should exceed $4.5$. In fact, the problem of bound states in a plasma and their breakup due to quantum effects and screening (Mott effect) has been discussed in many text books, see e.g. Ref.~\cite{kremp_05} and references therein. Recent first principle path integral Monte Carlo simulations \cite{bonitz_prl_05} indicate that bound states vanish in the density range around $r_s=1.5\dots 2$.  

While in our paper, the dynamically screened potential has been considered for the case of protons, it is trivial to apply these results directly to highly charged ions with $Q_{i}=Ze_0$. In this case the shape of the ion potential will remain the same, only its magnitude increases by a factor $Z$ compared to the present results. The same scaling applies to the perturbation of the electron density.

Finally, the dynamically screened effective potential approach can be directly used for MD simulations of classical ions on the background of streaming quantum electrons as discussed in Ref.~\cite{Patrick2}. This allows to obtain first principle static and dynamic results for the ion component, including the range of strong ion coupling. 
At very high densities, when the ion de Broglie wave length approaches the mean distance between ions, ionic quantum effects have to be taken into account \cite{ion-pseudo-pot}. At moderate ion degeneracy, this can be done approximately by replcacing the interaction between ions by a quasi-classical potential using an idea due to Kelbg \cite{Kelbg}. In the mean time improved quantum potentials have become available, e.g. \cite{Filinov1,Filinov,Deutsch,Moldabekov}. Using this idea it will be an interesting task to derive an effective quantum potential that combines dynamical screening (at large distances) and quantum effects (at short distances), as suggested in Refs.~\cite{Patrick2,Baimbetov} and to extend this idea to potentials including long-range wake effects.

\subsection*{Acknowledgments}

We thank I. Schnell and H. K\"ahlert (Kiel) for helpful discussions and M. Thoma (Giessen) for bringing to our attention the related results for quark-gluon plasmas, Ref.~\cite{Toma}.
This work has been supported by the Deutsche Forschungsgemeinschaft via grant SFB-TR 24 TPA 9 and  the Ministry of Education and Science of
Kazakhstan under Grant No. 1415/GF2 2014 (IPC-21).

\section*{Appendix A: Finite temperature RPA dielectric function}\label{s:rpa}
Here we give a brief summary on the main formulas and the numerical implementation of the random phase approximation (RPA) dielectric function $\epsilon_{\rm RPA}$ for electrons in thermodynamic equilibrium at a finite temperature $T$ which is required to compute the dynamically screened potential of the ions, Eq.~(\ref{POT}). The retarded RPA-DF is given by
\begin{equation} \label{ RPAdk }
\epsilon_{\rm RPA}(\vec{k},\omega) = 1 - \frac{e^2}{\pi^2 k^2}\int d\vec{k}\sp{\prime} \frac{f(\vec{k}+\vec{k}\sp{\prime}) - f(\vec{k}\sp{\prime})}{E(\vec{k}+\vec{k}\sp{\prime} ) - E(\vec{k}\sp{\prime}) - \hbar{\hat \omega}},
\end{equation}
where ${\hat \omega}=\omega + i \delta$ ($\delta \to +0$), the electron single-particle energy is $E(\vec{k})=\hbar^{2}k^{2}/2m$, and $f(\vec{k})$ is the Fermi-Dirac distribution. Using the standard 
normalization to the density, $2\int d^3k \, f({\bf k}; T, \mu) = n$, for the spin-unpolarized case
\begin{equation}
 \frac{2}{3}\theta^{-3/2}=\int_0^\infty \frac{\sqrt{x}}{1+\exp(x-\beta \mu)} dx,
\nonumber
\end{equation}
allows for an inversion to yield the chemical potential as a function of density and temperature, $\beta\mu(\Theta)$.
%
%
\begin{figure}[h]
\vspace*{0.5cm}
\includegraphics[width=1\linewidth]{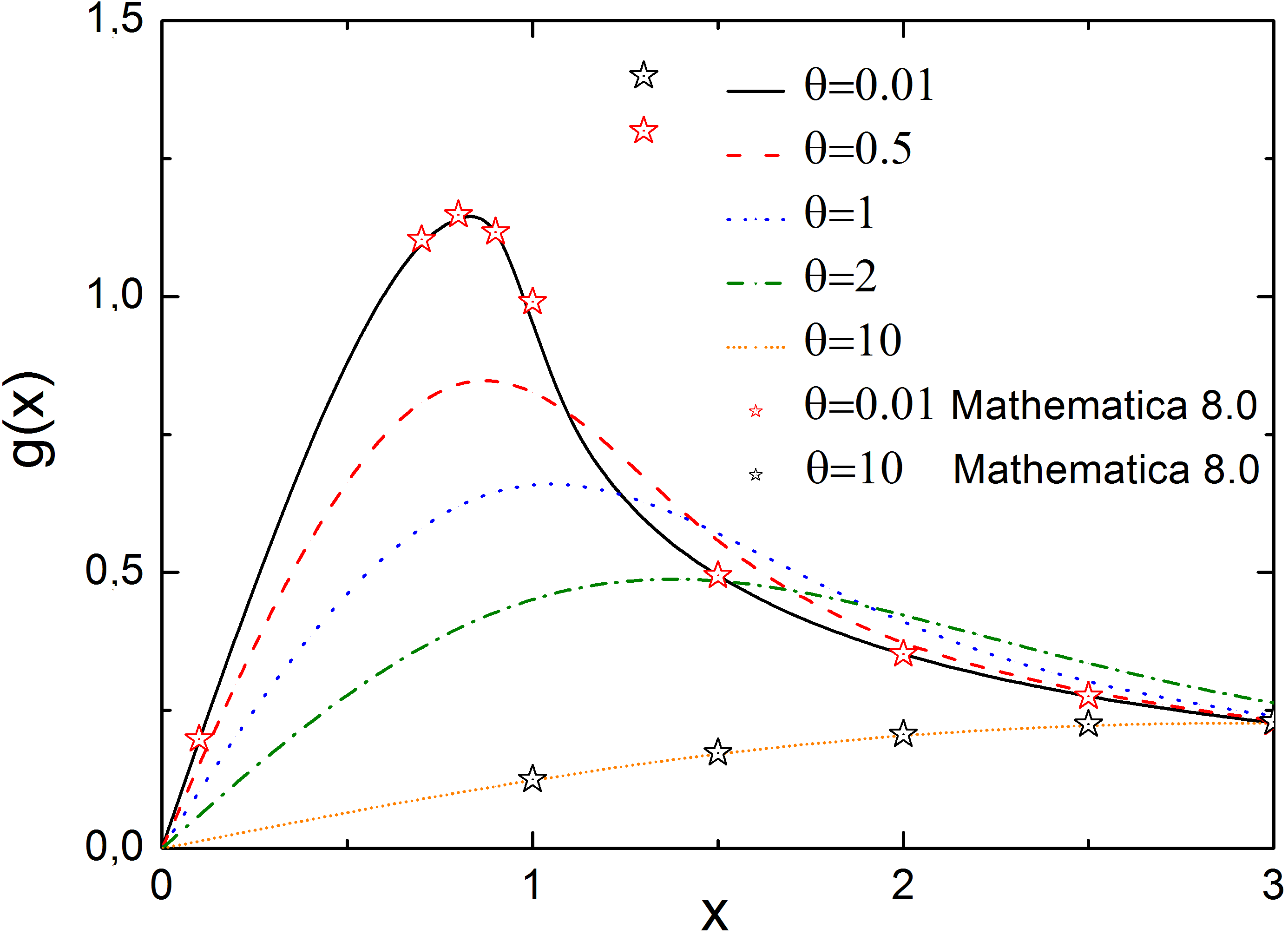}
\caption{Function $g(x)$ defined by Eq.~(\ref{ g }) for different values of $\theta$. Comparison of results of numerical integration and the implementation in Mathematica (symbols).}
\label{fig:gx}
\end{figure}
\begin{figure}[h]
\includegraphics[width=1\linewidth]{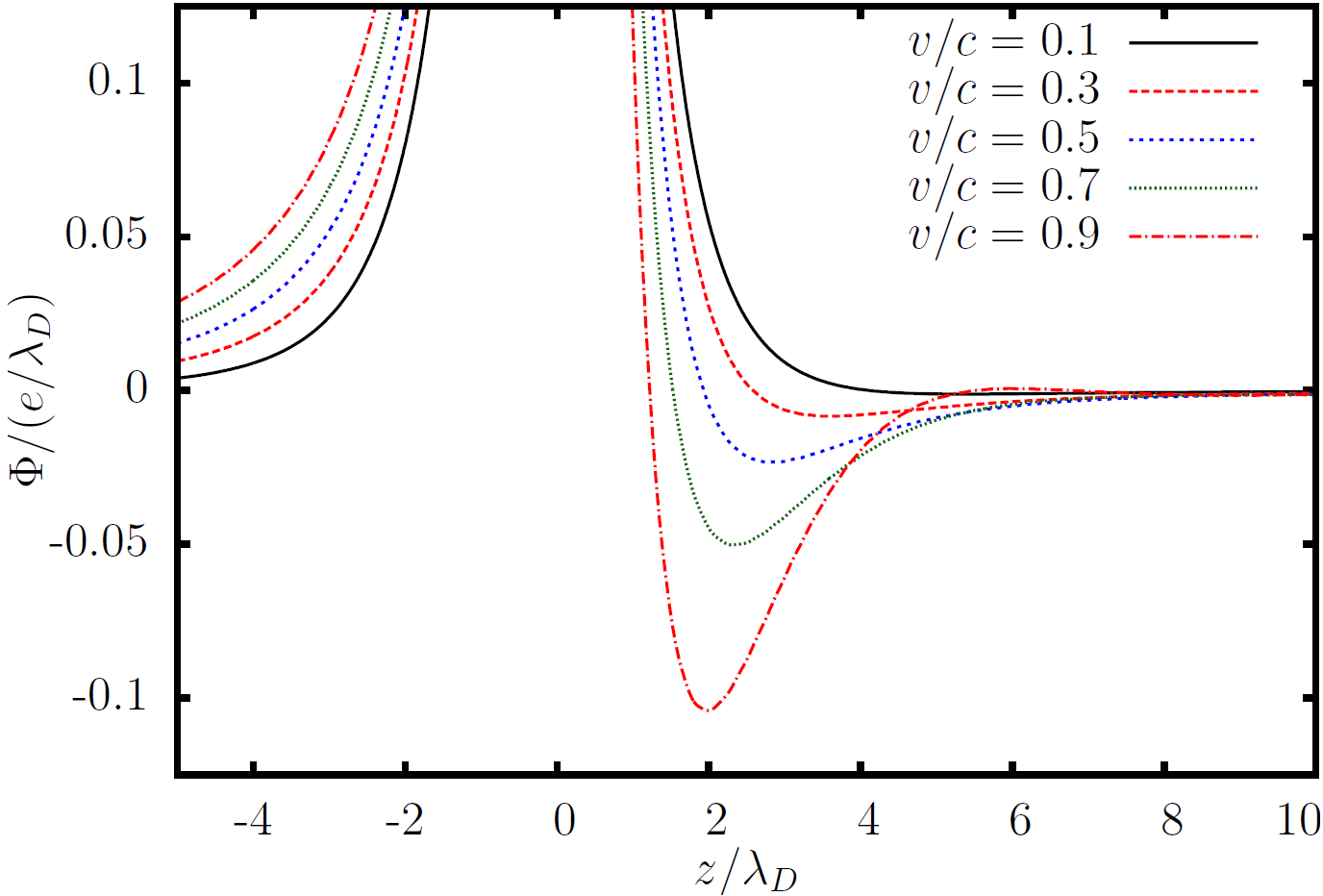}~\hspace{-0mm}\begin{scriptsize}a)\end{scriptsize}
\includegraphics[width=1\linewidth]{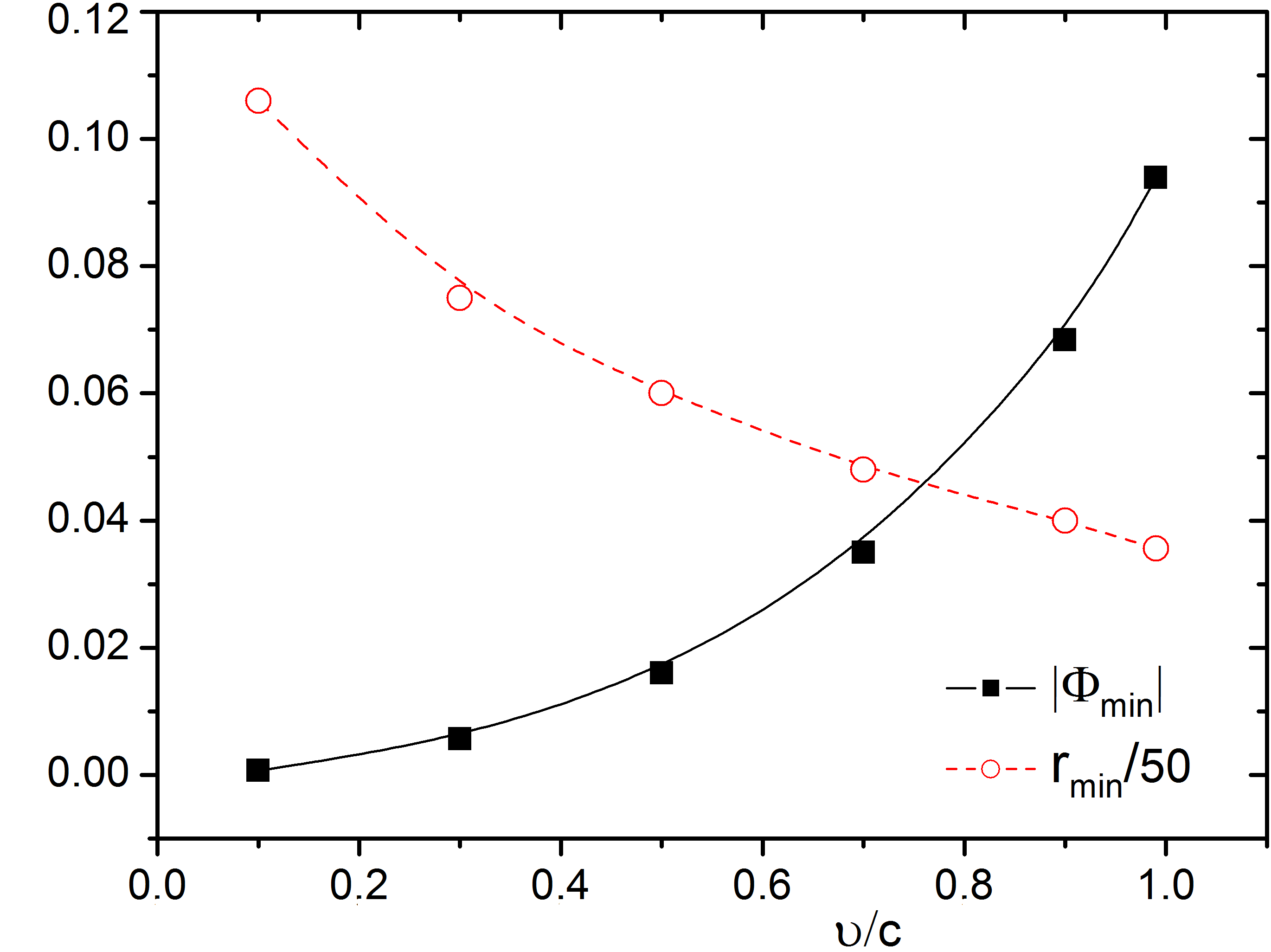}~\hspace{-0mm}\begin{scriptsize}b)\end{scriptsize}
\caption{Dynamically screened potential $\Phi(z)$ of an ultra-relativistic plasma in streaming direction. {\bf Top}: Potential for different values $M=v/c$. {\bf Bottom:} absolute depth and location  of the first minimum versus streaming parameter $M=v/c$.}
\label{fig:EPaxial}
\end{figure}
Separating the real and imaginary part of the dielectric function~(\ref{ RPAdk }),
\begin{equation} \label { eps }
 \epsilon_{\rm RPA}(k,\omega)=\epsilon_1(k,\omega)+i\epsilon_2(k,\omega),
\end{equation}
one obtains, e.g. \cite{A and B},
\begin{align} \label{dk1}
 \epsilon_1(k,\omega) &= 1+\frac{\chi^{2}}{4z^3}[g(u+z)-g(u-z)], 
\\
\epsilon_2(k,\omega) &= \frac{\pi\chi^{2}}{8z^3}\theta \ln \left(\frac{1+\exp(\beta\mu-(u-z)^2/\theta)}{1+\exp(\beta\mu-(u+z)^2/\theta)}\right),
\end{align}
where $u=\omega/kv_{F}$, $z=k/2k_{F}$, $\chi^{2}=1/\pi k_{F}a_{B}$, and $k_{F}$ is the Fermi momentum. The function
$g(x)$ depends parametrically on the degeneracy parameter $\theta$ [this dependence is suppressed in Eq.~(\ref{dk1})], 
\begin{equation} \label{ g }
 g(x)=\int_0^\infty \frac{ydy}{\exp\{y^2/\theta-\beta \mu\}+1}\ln \left| \frac{x+y}{x-y}\right|. 
\end{equation}
The integral in Eq.~(\ref{ g }) can be simplified in the limits of high and low degeneracy  \cite{A and B}.
However, a direct numerical integration poses no problem. A convenient and sufficently accurate implementation 
is available in Mathematica \cite{Mathematica}. Both results are shown in Fig.~\ref{fig:gx}.
\section*{Appendix B: Dynamically screened potential of an ultra-relativistic quantum plasma}\label{s:EP}
The goal of this appendix is to compare the results for the dynamically screened potential produced by degenerate non-relativistic electrons to the regime of an ultra-relativistic plasma and compare its shape to the results for the former. The first example is the quark-gluon plasma (QGP) that is expected to have existed immediately after the big bang  and has been produced in heavy ion collisions at RHIC and CERN. The second example is the ultra-relativistic electron-positron plasma (EPP) that is expected to be produced in supernova explosions or in magnetars and should become experimentally accessible with next generation high-intensity lasers. Without going into details (for a recent review and further references, see Ref.~\cite{Toma4}), we note that the common condition for these systems is that the temperature exceeds the rest mass of the particles, $T \gg mc^2$. Furthermore, for the dielectric analysis below it is assumed that the system is close to equlibrium and nearly ideal.

Since the dielectric functions of the QGP and EPP are, to lowest order, identical (except for simple factors related to the number of quark flavors),
we will concentrate, in the following on the EPP case, using the natural units, $\hbar=c=k_{B}=1$, common in quantum field theory.
The relativistic quantum dielectric function was derived by Silin \cite{silin_jetp60}, and simplified dispersions were obtained by many authors, see e.g. Ref.~\cite{braaten_prd93} and references therein.
An analytical formula for the ultra-relativistic longitudinal dielectric function that takes into account collisions which is analogous to the Mermin approximation of the present paper was derived by Thoma et.al \cite{Toma3} and has the form:
\begin{equation}\label{EPeps}
\begin{split}
\epsilon(k,\omega)&=1 +\frac{3m_{\gamma}^{2}}{k^{2}}\left(1-\frac{\omega+ i\nu}{2k}\ln\frac{\omega+i\nu+k}{\omega++i\nu-k}\right)
\\
& \cdot \left(1-\frac{i\nu}{2k}\ln\frac{\omega++i\nu+k}{\omega+ i\nu-k}\right)^{-1},
\end{split}
\end{equation}
where $m_{\gamma}=eT/3$ is the effective photon mass.

Using this result we now compute the real-space dynamically screened potential acording to Eq.~(\ref{POT}), using the computer code of the main part of the paper. 
For the electron-positron collision frequency we use $\nu/\omega_{p}=5.4\cdot10^{-4}$~\cite{Toma4}. In the figures below we show
 the effective potential in units of $e/\lambda_D$, where $\lambda_{D}=1/\sqrt{3}m_{\gamma}$ is the Debye screening length, whereas the dimensionless streaming velocity (Mach number) is defined as $M=u/c$.

In Fig.~\ref{fig:EPaxial}.a) the dynamically screened potential in streaming direction is shown for different streaming velocities. As in the non-relativistic case there exists an attractive minimum, the depth and position of which are plotted in Fig.~\ref{fig:EPaxial}.b). One clearly sees that the minimum becomes deeper and its location closer to the projectile when its velocity increases, as in the nonrelativistic case.
\begin{widetext}
The shape of the potential away from the symmetry axis is shown in Fig.~\ref{fig:EPshape}. As in the non-relativistic case [cf. Figs.~\ref{fig:wake}--\ref{fig:wake_M} ], the potential is ``bent''forward towards the projectile which confirms our discussion of the main text and the explanation in terms of quantum effects. A more detailed comparison of the non-relativistic and ultra-relativistic cases will be given elsewhere \cite{ludwig_cpp14}. 

\begin{figure}[h]
\includegraphics[width=0.3\textwidth]{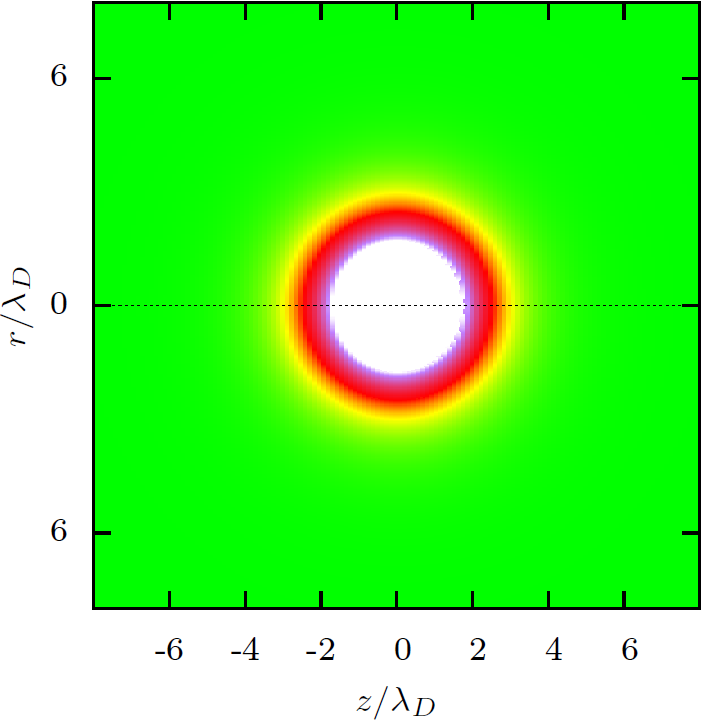}~\hspace{-4.7mm}
a)
\includegraphics[width=0.261\textwidth]{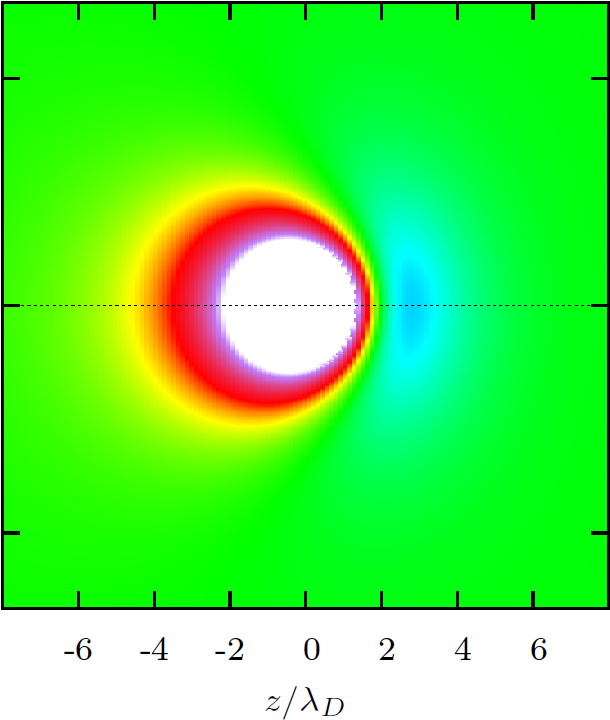}~\hspace{-4.3mm}
b)
\includegraphics[width=0.328\textwidth]{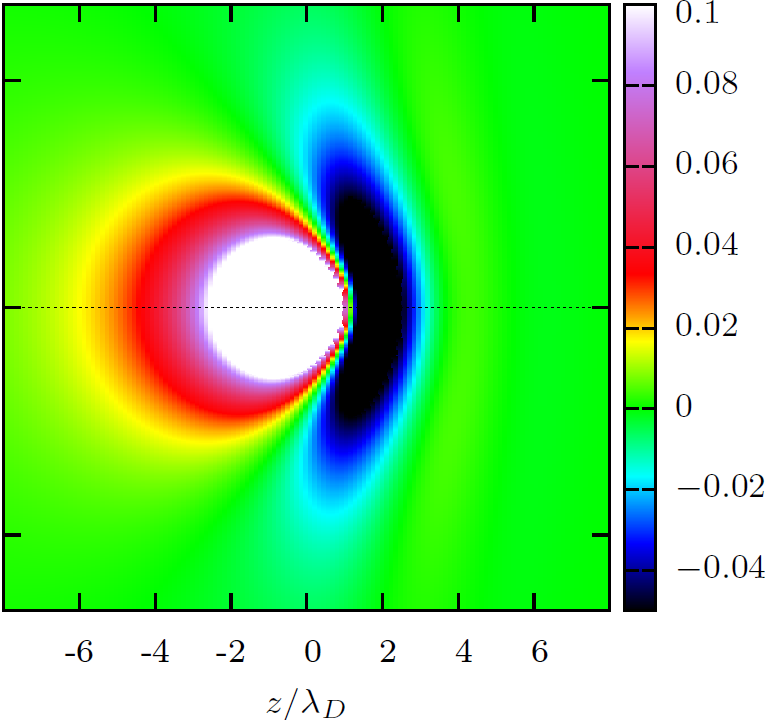}~\hspace{-18mm}
c)
\caption{(Color online) Dynamically screened potential $\Phi(r,z)/(e/\lambda_{D})$ of an ultra-relativistic plasma for three different streaming velocities $M=v/c$:  a) $M=0 $, b) $M=0.55 $, and c) $M=0.99 $.}
\label{fig:EPshape}
\end{figure}
\end{widetext}

\begin{thebibliography}{10}
\bibitem{df_damping} It is well known that going beyond  RPA by  including the collisions cannot be realized simply by substituting $\omega$  by $\omega+i\nu$. This method fails to conserve the local particle number and does not have the correct asymptotics.
\end{thebibliography}
\end{document}